\documentclass[%
reprint,
superscriptaddress,
nofootinbib,
amsmath,amssymb,
aps,
floatfix,
]{revtex4-2}

\usepackage{graphicx}
\graphicspath{ {figures/} } 

\usepackage{dcolumn}
\usepackage{bm}
\usepackage{booktabs}  
\usepackage{csquotes}
\usepackage{siunitx}  
\usepackage{physics}
\usepackage{xcolor}
\usepackage{breqn}  

\usepackage{hyperref}

\usepackage{cleveref} 
\newcommand{\lsim}{\lesssim}

\newcommand{\eq}[1]{Eq.~(\ref{#1})}

\newcommand{\ord}[1]{\mathcal{O}{(#1)}}
\newcommand{\beq}{\begin{equation}}
\newcommand{\eeq}{\end{equation}}
\newcommand{\bea}{\begin{eqnarray}}
\newcommand{\eea}{\end{eqnarray}}

\newcommand{\appropto}{\mathrel{\vcenter{
  \offinterlineskip\halign{\hfil$##$\cr
    \propto\cr\noalign{\kern2pt}\sim\cr\noalign{\kern-2pt}}}}}

\newcommand{\upperlimit}{Irregularities of the low values are due to the
inability to determine the exact precession rate from the simulation results.
Hence, the points only show a statistical upper limit of the possible vertical
precession rate, actual rates could be lower. More about this in
\Cref{sec:data}.} 
\makeatletter
\preto\maketitle{%
  \begingroup\lccode`~=`,
  \lowercase{\endgroup
  \let\saved@breqn@active@comma~
  \let~}\active@comma 
}
\appto\maketitle{%
  \begingroup\lccode`~=`,
  \lowercase{\endgroup
  \let~}\saved@breqn@active@comma 
}
\makeatother
\begin{document}

\preprint{APS/123-QED}

\title{Comprehensive Symmetric-Hybrid ring design for a pEDM experiment at below
  $10^{-29}e\cdot \mathrm{cm}$}

\author{Zhanibek Omarov}
\affiliation{Department of Physics, KAIST, Daejeon 34141, Republic of Korea}
\affiliation{Center for Axion and Precision Physics Research, IBS, Daejeon 34051, Republic of Korea}

\author{Hooman Davoudiasl}
\affiliation{High Energy Theory Group, Physics Department, Brookhaven National Laboratory, Upton, New York 11973, USA}

\author{Selcuk Hac\i\"omero\u glu}
\email{selcuk.haciomeroglu@gmail.com}
\thanks{Corresponding author}
\affiliation{Center for Axion and Precision Physics Research, IBS, Daejeon 34051, Republic of Korea}

\author{Valeri Lebedev}
\affiliation{Fermi National Accelerator Laboratory, Batavia, IL 60510, USA}

\author{William M. Morse}
\affiliation{Brookhaven National Laboratory, Upton, New York 11973, USA}

\author{Yannis K. Semertzidis}
\email{semertzidisy@gmail.com}
\thanks{Corresponding author}
\affiliation{Center for Axion and Precision Physics Research, IBS, Daejeon 34051, Republic of Korea}
\affiliation{Department of Physics, KAIST, Daejeon 34141, Republic of Korea}

\author{Alexander J. Silenko}
\affiliation{Bogoliubov Laboratory of Theoretical Physics, Joint Institute for
  Nuclear Research, Dubna 141980, Russia}
\affiliation{Institute of Modern Physics, Chinese Academy of Sciences, Lanzhou 730000, China}
\affiliation{Research Institute for Nuclear Problems, Belarusian State University, Minsk
220030, Belarus}

\author{Edward J. Stephenson}
\affiliation{CEEM, Indiana University, Bloomington, Indiana 47408, USA}

\author{Riad Suleiman}
\affiliation{Thomas Jefferson National Accelerator Facility, Newport News, VA
  23606, USA}

\date{May 2021}

\begin{abstract}
  A concise demonstrative summary of the Symmetric-Hybrid ring design
  for the storage ring proton electric dipole moment experiment is presented.
  Critical issues such as lattice design, background electrical fields,
  geometrical phase, general relativity, spin coherence time and polarimeter
  systematics are presented. Overall, we find that with the currently proposed
  design iteration, the systematic error sources are reduced by orders of magnitude
  and that the ring alignment requirements are within  currently available technology.
\end{abstract}

\maketitle

\section{Introduction}
The latest muon \((g-2)\) results~\cite{fnal1,fnal2,fnal3,fnal4} have
demonstrated the high sensitivity reach of experimental, analytical and simulation
tools with the latter matching and many times surpassing the precision of analytical estimations. Spin and beam dynamics needed to be understood with high precision
similar to the level required for a sensitive storage ring electric dipole moment (EDM)
experiment. This article describes a high precision storage ring EDM experiment
for the proton as the next generation of high precision and high impact physics
in storage rings.

The EDM of an elementary particle is proportional to its spin $\vec{S}$, which
is odd under time reversal $T$.  Hence, in the presence of an electric field
$\vec{E}$, which is invariant under $T$, the interaction Hamiltonian of the
particle $H_{\rm int}\propto -\vec{E}\cdot\vec{S}$ violates $T$ symmetry.  This
would also imply combined charge conjugation parity (CP) symmetry violation,
given CPT conservation, which is encoded in the quantum field theory formulation
of the Standard Model (SM).

The weak interactions in the SM mediate well-established CP violating phenomena
and can, through quantum processes, generate non-zero EDMs for constituents of
atoms, {\it i.e.} electrons and nucleons.  However, the electron and nucleon
EDMs in the SM are induced at high loop orders and are quite suppressed:
$d_e^{\rm SM} \lsim 10^{-38}$ $e\cdot \mathrm{cm}$ and $d_N^{\rm SM}\lsim 10^{-32}$ $e\cdot \mathrm{cm}$,
respectively; $N=p, n$ \cite{Pospelov:2005pr,Engel:2013lsa,Chupp:2017rkp}.  The
EDMs generated by the SM interactions are not observable at current or
near-future experiments, making any positive measurement an unambiguous signal
of new physics.

It is interesting to note that the SM, in principle, could have generated a
large nucleon EDM, through a $P$-odd and $T$-odd renormalizeable interaction
$\propto \theta \,G_{\mu\nu} \tilde{G}^{\mu\nu}$, where $\theta$ is a
fundamental parameter of quantum chromodynamics (QCD); $G_{\mu\nu}$ and
$\tilde{G}^{\mu\nu}$ denote the field strength tensor and dual tensor of the
gluon, respectively.  Due to the axial anomaly of QCD, the value of
$\theta$ gets shifted when quarks are transformed by chiral rotations that
diagonalize the quark mass matrix $M_q$.  Thus, the physically observable
quantity is given by
\beq
\bar \theta \equiv \theta + \arg[\det(M_q)].
\label{thetabar}
\eeq

The contribution of $\bar \theta$ -- assuming dominance of the long-range pion
loop processes -- to nucleon EDMs is estimated to be
\cite{Crewther:1979pi,Pospelov:2005pr,Mereghetti:2010kp,Dragos:2019oxn} ($q=e>0$ is the charge of the proton),
\beq
-d_n(\bar \theta) \approx d_p(\bar \theta) \approx 10^{-16} \bar{\theta}\, e\cdot \mathrm{cm}.
\label{NEDM}
\eeq
However, the above relation does not in general hold, since there are short
range contributions to $d_n(\bar{\theta})$ and $d_p(\bar{\theta})$ that can in
principle have similar magnitudes as that in \eq{NEDM}.  There is no reason to
expect that the long- and short-range contributions should cancel, and hence one
can take the above estimate as a  good lower bound
\cite{Mereghetti:2010kp,Dragos:2019oxn}, though a non-perturbative treatment is
required for a more definitive result; see for example
Refs.~\cite{Dragos:2019oxn,Izubuchi:2020ngl,bhattacharya2021contribution}.  Given the
current bound on the neutron EDM $d_n < 1.8\times 10^{-26}$~$e \cdot \mathrm{cm}$ (90\% C.L.) \cite{neutronEDM2020}, one then obtains $\bar \theta \lsim 10^{-10}$.

Note that $\bar\theta$ could be rotated away if one of the quarks is massless,
rendering $\arg[\det(M_q)]$ ill-defined.  That possibility is disfavored by low
energy hadron phenomenology and lattice computations
\cite{Drury:2013sfa,Bazavov:2018omf,Alexandrou:2020bkd}.  The smallness of
$\bar\theta$ is therefore a conceptual SM puzzle, since there is no obvious
reason why the sum of the contributions in \eq{thetabar} should cancel so
precisely.  A well-known resolution of this ``strong CP problem" is furnished by
the Peccei-Quinn (PQ) mechanism \cite{Peccei:1977hh,Peccei:1977ur} which
provides a dynamical relaxation of $\bar\theta$ to zero and gives rise to a
light pseudo-scalar, the ``axion" \cite{Weinberg:1977ma,Wilczek:1977pj}.
Nonetheless, contributions from new physics beyond the SM (BSM) can perturb the
PQ mechanism and induce a non-zero $\bar\theta$ \cite{Pospelov:2005pr}.

There are good reasons for assuming BSM phenomena (setting aside gravity which
is well-described by General Relativity).  A multitude of observations
\cite{Tanabashi:2018oca} have established that  $\sim 25\%$ of the cosmic energy
budget is made up of an unknown substance -- namely dark matter (DM) -- which
requires BSM physics ({\it e.g.}, the PQ axion which can be a good DM
candidate).  The visible Universe, which accounts for $\sim 5\%$ of the cosmic
total, has a dominance of ordinary matter over anti-matter, whose origin is an
open fundamental question.   In addition, the well-established flavor
oscillation of neutrinos calls for non-zero neutrino masses which, again, cannot
be accommodated in the minimal SM.  Taken together, one reaches the unavoidable
conclusion that BSM physics is required for a more complete description of
Nature.

Quite generically, BSM theories introduce new interactions with complex
couplings and, hence, additional sources of CP violation.  In fact, an amount
of CP violation well above the level that the SM provides is a requirement for a
successful mechanism to explain the cosmic dominance of matter, or equivalently,
the baryon asymmetry of the Universe \cite{Sakharov:1967dj}.  Therefore, new CP
violating physics and additional contributions to particle EDMs are motivated
from a number of key and empirically well-established facts about Nature, apart
from any conceptual or theoretical arguments.

The SM prediction for a nucleon EDM, while a challenging experimental target, is
only about three orders of magnitude below the projected reach of a proton
storage ring facility, $\sim 10^{-29}e\cdot \mathrm{cm}$.  Thus, such an experiment has
excellent prospects either to find evidence for new physics, or else severely
constrain it; we will elaborate on this point in the following.

Numerous BSM proposals have been put forth over the last few decades to address
the shortcomings of the SM.  Many of these ideas have aimed to address the
``hierarchy" between the weak scale $\sim 100$~GeV and much larger mass scales,
such as the Planck mass $M_{\rm Pl}\sim 10^{19}$~GeV associated with possible
quantum gravity effects.  Models based on supersymmetry, weak scale
compositeness, and extra dimensions are some well known examples.  Theories that
attempt to explain the hierarchy generally predict the emergence of new physics
at energy scales $\lsim $~TeV, providing promising targets for discoveries at
the LHC.  However, so far the experiments at the LHC have not yielded any
conclusive evidence for BSM physics at $\ord{\rm TeV}$ energies.

The above state of affairs has in part prompted discussions about future
accelerators that can probe well beyond the TeV scale.  The enormous cost of
such facilities makes it imperative to provide strong physics motivations for
their discovery prospects.  For example, a $pp$ collider at center of mass
energy $\sqrt{s} = 100$~TeV, based on current analyses \cite{Golling:2016gvc},
could potentially access new states up to masses $M_{\rm new}\sim 30$~TeV.
While one could speculate about various BSM scenarios that may be discovered at
that facility, the detection of a clear proton EDM signal could provide
extremely compelling motivation for its construction, as we will briefly discuss
below.

Using quark models of hadrons, nucleon EDMs are estimated to be similar in size
to quark EDMs and color EDMs which involve gluons instead of photons.  An
order-of-magnitude estimate for the 1-loop quark EDM is
\beq
d_q \sim \frac{g^2}{16 \pi^2} \frac{e \,m_q\, \sin \phi}{M_{\rm new}^2}\,,
\label{dq}
\eeq
where $g$ is a typical coupling of new physics to a quark with mass
$m_q \sim 5$~MeV and $\phi$ is a BSM CP violating phase.  A dipole operator
couples left- and right-handed fermions and requires a chiral flip, accounted
for by the $m_q$ dependence of the above expression.  Let us assume a
loop-factor $g^2/(16 \pi^2) \sim 0.01$, as a typical expectation.  We then find
\beq
d_q \sim 10^{-29} \left(\frac{30 \,\text{TeV}}{M_{\rm new}}\right)^2 \left(\frac{\sin\phi}{0.01}\right) e\cdot \mathrm{cm}.
\label{dq-num}
\eeq
Thus, under reasonable assumptions, an EDM signal at a proton storage ring
experiment can provide a strong physics case for the significant investments
required to access scales of $\ord{10~\text{TeV}}$ at a future collider.  We
also point out that $\sin\phi\sim 0.01$ can be considered fairly conservative,
given that for the CP violating phase $\delta$ in the SM quark sector
$\sin\delta \sim 1$ \cite{Tanabashi:2018oca}.  If we take $\sin\phi\sim 1$ in
the BSM sector also, scales up to $\sim 300$~TeV can possibly be probed through
the proton EDM measurement, well beyond the reach of any collider envisioned for
the foreseeable future.

The recent years have seen a surge of interest in new ideas for BSM particles at
or below the GeV scale that have  suppressed coupling to the SM; see, for
example, Refs.~\cite{Essig:2013lka,Battaglieri:2017aum}.  Such physics may
originate from a ``dark sector" that includes DM and only indirectly interacts
with the visible world.  Also, given the apparent absence of BSM states near the
TeV scale, it is worth considering that new physics could have a low energy
scale, but require intense sources to access, due to its feeble interactions
with the SM.  Adopting this point of view, one may consider
$M_{\rm new}\sim 1$~GeV in \eq{dq}, which yields a proton storage ring
sensitivity to $g\lsim 3\times 10^{-5}$.  This greatly exceeds current and
projected sensitivity for the coupling of new light states to quarks, under
various assumptions for BSM physics; see for example
Refs.~\cite{Tulin:2014tya,Dror:2017ehi,Ilten:2018crw}.

Finally, we note that the first results of the experiment E989 at Fermilab that
were released recently \cite{fnal1} confirm the long-standing muon $g-2$
measurements at Brookhaven National Laboratory \cite{bennett_final_2006}.  The
combined results point to a 4.2$\sigma$ deviation from the SM theory prediction
of Ref.~\cite{Aoyama:2020ynm} (however, see also Ref.~\cite{Borsanyi:2020mff}).
If this deviation persists with more data and further scrutiny of the SM theory,
it would be a harbinger of new physics.  That physics could potentially also
manifest itself via a proton EDM measurement.  In that case, the complementary
precision signals from the lepton and hadron sectors could provide valuable
insights about the nature of the underlying BSM phenomena and help chart a
course for a new era of discovery.

The storage ring proton EDM method targets $d_{p}=10^{-29}e \cdot \mathrm{cm}$ which is
more than 3 orders of magnitude better than the current best neutron EDM
limits~\cite{neutronEDM2020}. We also claim that this sensitivity is achievable
with existing technology thanks to the significantly relaxed alignment
requirements with the Symmetric-Hybrid ring design.

There are multiple ways to design a lattice capable of measuring a charged
particle EDM, some of which are described in \Cref{tab:lattices}. Although there
are a number of choices, the one with the least systematic error sources
(potential risk) is chosen here for a comprehensive study --- Symmetric-Hybrid
design.

Although the direct measurement of charged particle EDM is challenging, the Muon $(g-2)$ experiments using storage rings have been setting the best direct EDM limits on muons. Similar to
$(g-2)$, the proton EDM also uses the so-called ``magic momentum''; though, the muon $(g-2)$ experiment at Fermilab uses magnetic bending and electric focusing to study the muon magnetic anomaly with high precision, the proton EDM proposal is to use electric bending and alternate magnetic focusing as the best way to reduce systematic error sources.
 No magnetic bending leads to lock-in of the average spin
directions with the momentum which is also recognized as ``frozen'' spin (more
in \Cref{sec:technique}).  A non-zero EDM causes a linear vertical spin build-up that is
measured as a function of beam storage time to infer the EDM value.

\begin{table*}
  \centering
  \caption{Brief description of storage ring designs capable of measuring charged particle
EDMs.}\label{tab:lattices}
  \begin{tabular}[t]{p{0.35\linewidth} @{\hskip 0.5in} p{0.6\linewidth}}
    \toprule
    Lattice & Comments \\ \midrule

    Muon $(g-2)$
            & Tipping angle of the $(g-2)$ precession plane~\cite{bnl_edm} lets us infer the muon EDM
              value. Limited statistical EDM sensitivity. When electric focusing is used, eventually it will be limited by geometrical alignment, which could require consecutive clockwise (CW) and
              counter-clockwise (CCW) injections to eliminate it. \vspace{0.5cm}\\

    $E,B$ fields combined lattice for measuring EDMs of Deuteron~\cite{ags_proposal},
    $^3\textnormal{He}$, proton, etc.

            & Right combination of $E,B$ fields leads to ``frozen-spin''
              condition, in principle, at any energy. High statistical sensitivity on EDM. Requires
              consecutive CW and CCW injection with flipping of the $B$-fields
              to eliminate the main systematic error source --- background
              vertical electric field (assuming magnetic quadrupoles). 
              Need to demonstrate strict stability of $E$-field direction with magnetic field flips. \vspace{0.5cm}\\

    All-electric (4-fold)~\cite{anastassopoulos_storage_2016}. Electric bending
    and weak vertical electric focusing.
            & Requires state-of-the-art magnetic shielding and ability to observe
            vertical separation of counter rotating beams at below \si{\nm} level.
            \vspace{0.5cm}\\

    Hybrid (4-fold)~\cite{haciomeroglu_magnetic_2019}. Electric bending and
    alternate (strong) magnetic focusing.
            & Does not require magnetic shielding due to the effective shielding from radial
              magnetic fields via magnetic quadrupoles. The lattice is still sensitive
              to vertical velocity systematic error source (\Cref{sec:vertical_velocity}) affecting mainly 
              the DM/DE sensitivity\footnote{DM/DE sensitivity refers to vertical spin build-up due 
              to unavoidable radial spin component (see \Cref{sec:vertical_velocity}), DM/DE and EDM separation is discussed in Ref. 
              \cite{graham_paper}. Importantly, a finite EDM is not a necessary condition for the DM/DE experiment and the other way around.} with still significant but manageable impact on EDM measurement. \vspace{0.5cm}\\

    Symmetric-Hybrid (this work). Same as Hybrid (4-fold) with maximal possible
    symmetry.
            & Including the benefits of the Hybrid (4-fold) ring, this design effectively
              eliminates the largest systematic error source --- vertical
              velocity (\Cref{sec:vertical_velocity}), impacting mostly the  DM/DE sensitivity. It also makes the EDM experiment easier by reducing this potentially large systematic error source. \\

    \bottomrule
\end{tabular}
\end{table*}

The most prominent systematic error source in the storage ring designs based on
the All-electric ring~\cite{anastassopoulos_storage_2016} is the background
radial magnetic field --- $B_{x}^{\textnormal{external}}$. The stray magnetic field
is the most challenging requirement~\cite{haciomeroglu_magnetic_2019}.
To overcome such a shielding requirement, the next iteration after the
All-electric ring, the Hybrid (4-fold) ring design~\cite{haciomeroglu_hybrid_2018} was
developed. It has been a major accomplishment since any
$B_{x}^{\textnormal{external}}$ is naturally shielded by the magnetic focusing
system. The Hybrid (4-fold) ring design features a strong alternating magnetic focusing with
electric bending that still allows simultaneous Clockwise (CW) and
Counter-Clockwise (CCW) beam storage. Counter-rotating (CR) beams are crucial to
avoid the first order systematic error source --- a vertical dipole $E$ field.

In rings where the main vertical focusing is magnetic (e.g. Hybrid (4-fold)),
the main systematic error source becomes the out-of-plane (vertical) electric
field. However, this systematic error cancels exactly for vertical dipole
electric fields with CR beams. It is the only lattice that accomplishes this
cancellation and as such it represents a major breakthrough in the storage ring
EDM field. The next level systematic error source is the fact that the average
vertical velocity integrated over electric field sections might not be zero.
This is a strict requirement for the case of radial polarization, able to probe 
Dark Matter and Dark Energy (DM/DE) \cite{graham_paper}, and has been
relaxed by several orders of magnitude by making the lattice highly symmetric.

In this work, the newest design iteration, the Symmetric-Hybrid ring relaxes
requirements established by the Hybrid (4-fold) ring by several orders of magnitude,
provides comprehensive systematic error analysis, and standardizes experimental
techniques. Highlighted novelties of this work include,
\begin{itemize}
  \item Symmetric-Hybrid lattice design (\Cref{sec:technique,sec:spinbasedalignment})
  \item Spin-based alignment (\Cref{sec:spinbasedalignment})
  \item Hybrid sextupole configuration for simultaneous spin coherence time
        (SCT) improvement for CR beams (\Cref{sec:sct}).
\end{itemize}
By providing solutions to the most significant systematic error
sources and designing a storage ring with realistic specifications,
this work aims to be the foundational basis for the storage ring proton EDM
experiment. 

The rest of the paper is structured as follows, \Cref{sec:methods} provides an
introduction to the experimental technique and the tools used in this work,
\Cref{sec:systematicerrors} discusses the major systematic error sources, and
\Cref{sec:discussion} concludes the work by providing the relevant discussions.

\section{Methods}\label{sec:methods}
\subsection{Experimental Technique}\label{sec:technique}
The spin $\vec{S}$ precession rate for a particle at rest in the presence of magnetic
$\vec{B}$ and electric $\vec{E}$ fields is given as,
\begin{equation}
  \frac{d\vec{S}}{dt} = \vec{\mu}\cross\vec{B} + \vec{d} \cross \vec{E}
  \nonumber
\end{equation}
where magnetic and electric dipole moments are defined as
$\vec{\mu} = (gq/2m)\vec{S}$ and $\vec{d} = (\eta q/2mc)\vec{S}$ respectively.

Spin motion relative to the momentum for a particle with
$\vec{\beta} = \vec{v}/c$ in a cylindrical coordinate system\footnote{i.e. in
standard right-handed accelerator Frenet-Serret $x,y,s$ coordinates. This
coordinate system is used throughout the work unless stated otherwise.}, is given
as~\cite{omega1,bmt2,bmt3},
\begin{dmath*}[breakdepth={3}]
  \vec{\omega}_{a}  =  -\frac{q}{m} \left(G\vec{B}
    - \frac{G\gamma}{\gamma+1} \vec{\beta} (\vec{\beta}\cdot\vec{B})
    - \qty(G-\frac{1}{\gamma^{2}-1}) \frac{\vec{\beta}\cross\vec{E}}{c}
    + \frac{1}{\gamma} \left[\vec{B}_{\parallel} - \frac{1}{c\beta^{2}}
      \left(\vec{\beta}\times\vec{E}\right)_{\parallel}\right]
  \right)
\end{dmath*}
\[
  \vec{\omega}_{\eta} = -\frac{\eta q}{2m} \left( \frac{\vec{E}}{c} - \frac{\gamma}{\gamma+1} \frac{\vec{\beta}}{c} \qty(\vec{\beta} \cdot \vec{E}) + \vec{\beta}\cross\vec{B} \right),
\]
where $\omega_{a}$ and $\omega_{\eta}$ stand for precession due to magnetic and electric dipole
moments respectively, $G$ stands for the proton magnetic anomaly. For $\vec{\beta} \cdot \vec{E} = 0$ and $ \vec{\beta}\cdot\vec{B}=0$
the motion of the spin vector simplifies more,
\begin{dmath}
  \vec{\omega}_{a}  =  -\frac{q}{m} \left(G\vec{B} - \qty(G-\frac{1}{\gamma^{2}-1})\frac{\vec{\beta}\cross\vec{E}}{c} + \frac{1}{\gamma}\left[\vec{B}_{\parallel}-\frac{1}{c\beta^{2}}\left(\vec{\beta}\times\vec{E}\right)_{\parallel}\right]\right)
  \label{eq:omega_a}
\end{dmath}
\begin{equation}
  \begin{split}
    \vec{\omega}_{\eta} & = -\frac{\eta q}{2m} \qty(\frac{\vec{E}}{c} + \vec{\beta}\cross\vec{B}) \\
    \vec{\Omega} & = \vec{\omega}_{a} + \vec{\omega}_{\eta}
  \end{split}
  \nonumber
\end{equation}
\begin{equation}
  \frac{d\vec{S}}{dt} = \vec{\Omega} \times \vec{S},
  \label{eq:dsdt}
\end{equation}
with $\parallel$ indicating horizontal (in-plane) projection of a vector.

We set $\vec{B} = 0$\footnote{Setting $\vec{B}=0$ is not technically correct due
to having magnetic quadrupoles, but it is helpful to assume temporarily.} and choose ``magic momentum'' such that
$\gamma = \sqrt{1 + 1/G}$. For protons, the ``magic'' parameters are given on
\Cref{tab:magic_momentum}. By choosing such proton momentum,
\Cref{eq:omega_a} leads to
\begin{equation}
 \vec{\omega}_{a} = \frac{q}{m\gamma c\beta^{2}} \left(\vec{\beta}\times\vec{E}\right)_{\parallel}.
  \label{eq:dsdt2}
\end{equation}
Notably, a vertical electric field would create
a non-zero radial component for $\vec{\omega}_{a}$, which would look like the
EDM signal with one beam direction. With horizontal $\vec{E}$ fields and $\vec{\beta}$ ($\vec{E}=\vec{E}_{\parallel}$,
$\vec{\beta}=\vec{\beta}_{\parallel}$), the
equation simplifies further into $\vec{\omega}_{a} = 0$ which is also known as
the frozen spin condition.  In this arrangement, the spin precesses into
the vertical direction only due to the EDM contribution, \[\Omega \propto \eta {E}\]
linearly in the time scale of the injection
${\Omega} \propto dS_{y}/dt$.
$dS_{y}/dt \propto E \eta$ is the fundamental principle of measuring the proton EDM.
That is, measurement of the out-of-plane spin precession rate ($dS_{y}/dt$)
inside a storage ring probes the intrinsic EDM of the particles. The coupling of
the electromagnetic fields to a particle's magnetic dipole moment (MDM) is
orders of magnitude larger than the EDM coupling. Hence, a strict alignment requirement of
electromagnetic fields is necessary. Further details about the storage ring EDM
experiment could be found in Refs.~\cite{edm_proposal,farley_new_2004}.

\begin{figure}[tbp]
  \centering
  \includegraphics[width=.99\linewidth]{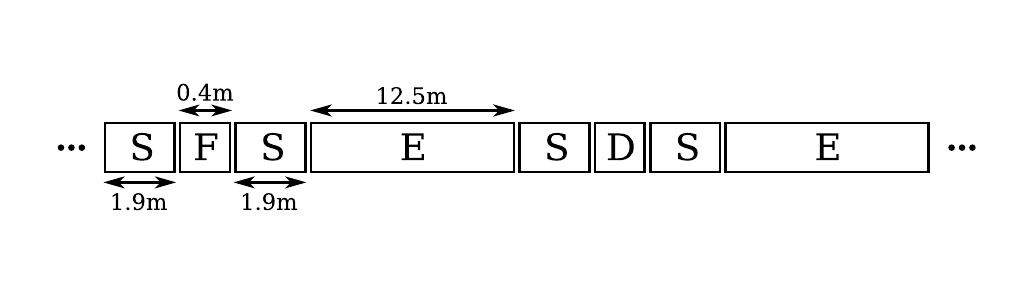}
  \caption{Schematic view of a single FODO cell. The entire ring is composed by
    stacking this unit 24 times. Legend: F --- Focusing
    quadrupole, D --- Defocusing quadrupole, S --- Straight free drift, E --- Electric bending.}\label{fig:FODO}
\end{figure}

\begin{figure}[tbp]
  \centering
  \includegraphics[width=.99\linewidth]{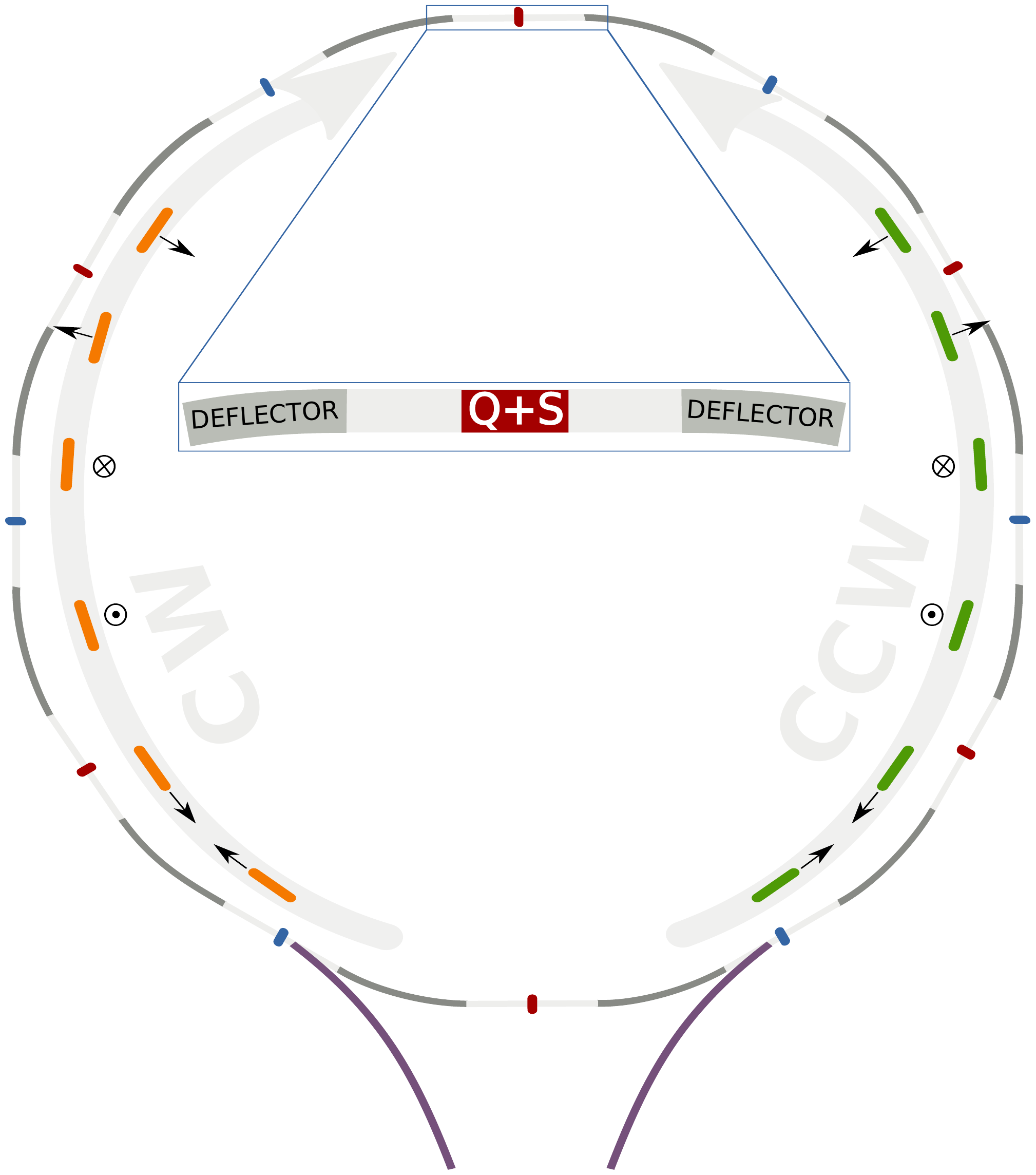}
  \caption{Schematic top view of the Symmetric-Hybrid ring. Both CR beams have
    longitudinally, radially, and vertically polarized bunches with different
    helicities (arrows in dark color). Blue and red correspond to focusing and defocusing quads. Naturally, CR beams see opposite focusing effect from magnetic quads. The actual number of FODO sections is 24.}\label{fig:ccw-cw}
\end{figure}

\begin{figure}[tbp]
  \centering
  \includegraphics[width=.99\linewidth]{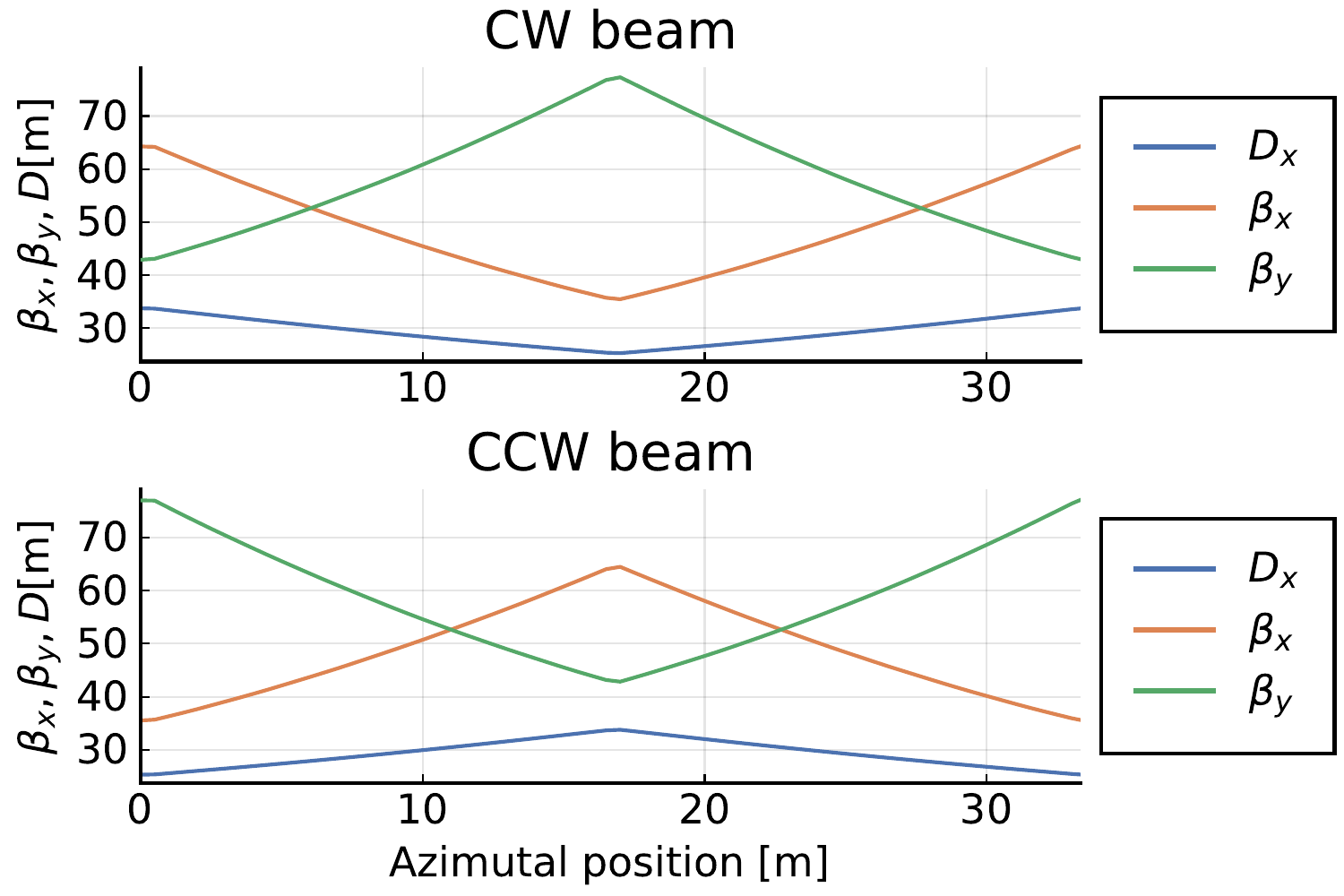}
  \caption{Superperiod structure, beta functions and dispersion ($\beta$ letter within text of the paper always refers to velocity).}\label{fig:beta}
\end{figure}

\begin{figure}[tbp]
  \centering
  \includegraphics[width=.99\linewidth]{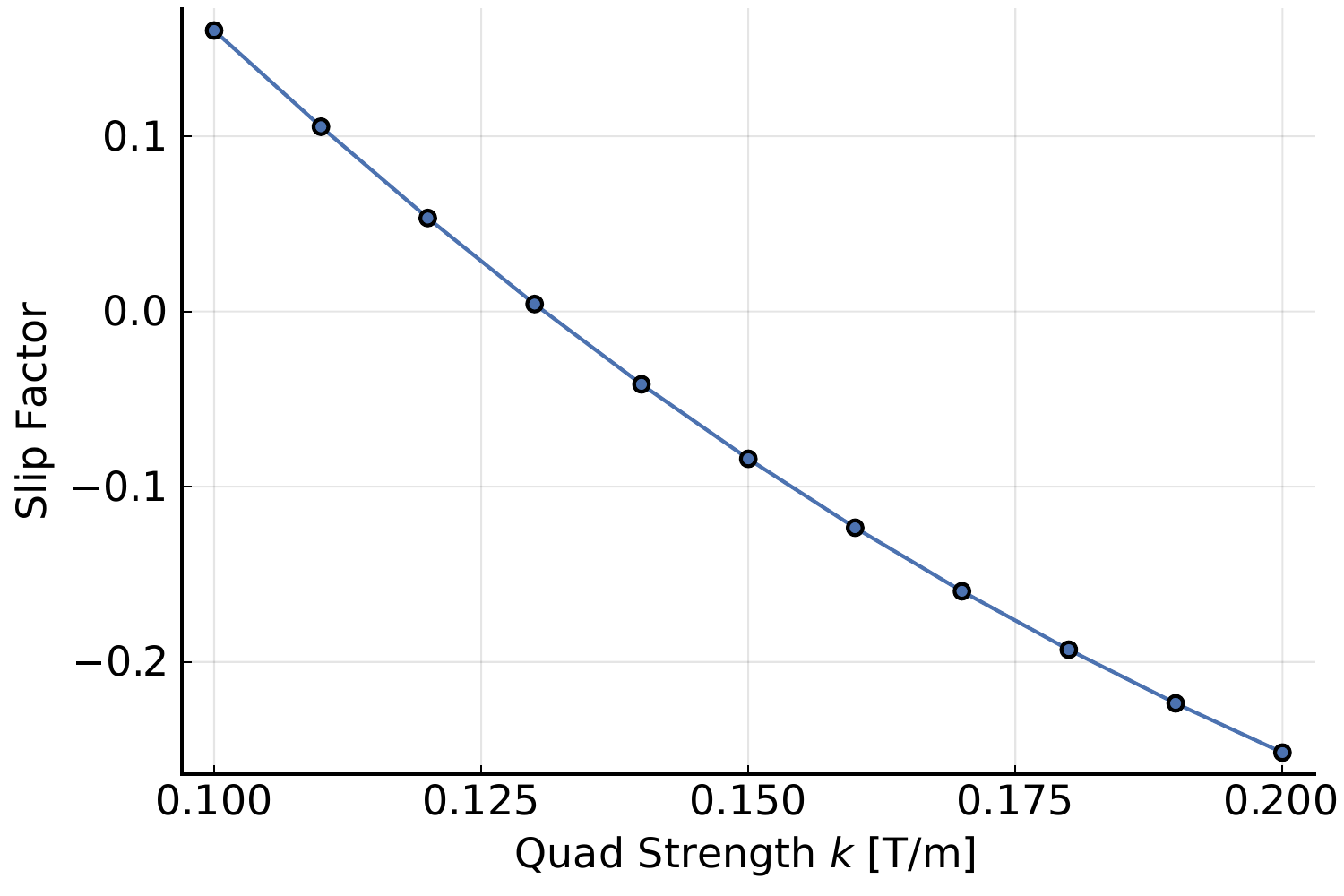}
  \caption{The slip factor is obtained from evaluating
    $\frac{dt}{t}/\frac{dp}{p}$ per turn via
    numerical tracking. Negative slip factor corresponds to below transition operation, which is essential with the Intra Beam Scattering considerations.}\label{fig:slip}
\end{figure}

The Symmetric-Hybrid ring design used in this study consists of 24 FODO sections
making up \SI{800}{m} in longitudinal length. Each FODO section comprises
a pair of electric bending sections (more about electric fields and electrode
design is in \Cref{sec:plates}) and a pair of magnetic quadrupoles. An illustration
of a single FODO is given in \Cref{fig:FODO}. A
schematic of the ring is given in \Cref{fig:ccw-cw}. Dispersion and beta
functions are given in \Cref{fig:beta} and the slip factor is given in \Cref{fig:slip}.

The design leaves $\SI{4.16}{m}$ of straight sections between electrostatic
bends. The straight sections are chosen to be sufficiently long for a vertical
injection of CR beams, polarimeters, RF-cavities and other apparatus. Notably,
 straight sections could be made longer at the cost of increasing the electric field
strength in the bending sections, i.e. by changing the ratio of circular /
bending lengths while retaining the total length of the lattice. The injected beam momentum is quite soft and an injection scheme has been worked out assuming only presently standard technology. The CR beams will be injected one after the other with their polarization in the vertical direction. The beam will then be let to de-bunch, with an RF-cavity re-bunching both CR beams with parameters as shown on \Cref{tab:specs}. Subsequently, a standard RF-solenoid will be used to create bunches with longitudinal polarizations (both helicities) and radially polarized bunches pointing inwards/outwards of the ring center.

As the estimations have shown, the slip factor (\Cref{fig:slip}) needs to be
negative in order for the Intra-Beam Scattering (IBS) not to cause severe beam lifetime
issues.  With beam storage times of $\approx\SI{e3}{\sec}=\SI{17}{min}$ in mind,
IBS becomes the primary mechanism of the emittance growth and consequently of
particle loss. For the current ring design, the beam lifetime is estimated to be
$\SI{22}{min}$ due to IBS and residual gas scattering assuming the vacuum level
$\SI{e-10}{Torr}$ of atomic hydrogen equivalent.  The beam will be lost
primarily on the polarimeter target due to IBS--induced exchange of horizontal
and vertical emittance.  The betatron tunes are optimized to avoid resonances up
to 8\textsuperscript{th} order inclusively, with the consideration of space
charge and beam-beam tune shifts. The selected tunes were confirmed to be free of beam
resonances with simulation. Additionally, the lattice is
compatible with stochastic cooling, which might be used to further prolong the spin
coherence time (\Cref{sec:sct}) and the beam lifetime.

More specifications and details are given on \Cref{tab:specs}.

\subsection{High precision tracking}
A Runge-Kutta family integrator (5th order, adaptive step size
\cite{tsitouras_rungekutta_2011}) was used in order to perform simulations
throughout this work. It was cross-checked by at least one independent effort for most of the shown studies. Both beam and spin dynamics are fully tracked numerically.
Particle beam dynamics are treated with perturbative expansion of the Lorentz
equation around the reference orbit in a Frenet-Serret coordinate system, with
the spin tracked via the T-BMT equation. More details are given in \Cref{sec:tracking}.
\begin{table}[tbp]
  \centering
  \caption{``Magic'' parameters for protons, values obtained from Ref.~\cite{mooser_direct_2014}.}
  \begin{tabular}{lllll}
    \toprule
    $G$           & $\beta$         & $\gamma$        & $p$                  & $KE$
    \\
    \midrule
    $\num{1.793}\qq{}$ & $\num{0.598}\qq{}$ & $\num{1.248}\qq{}$ & $\SI{0.7}{GeV/c}\qq{}$ & $\SI{233}{MeV}$\\
    \bottomrule
  \end{tabular}
  \label{tab:magic_momentum}
\end{table}
\begin{table}[tbp]
  \centering
  \caption{Ring and beam parameters for the Symmetric-Hybrid ring design}
  \begin{tabular}[t]{lc}
    \toprule
    Quantity                                                      & Value                        \\
    \midrule
    Bending Radius $R_{0}$                                        & \SI{95.49}{m}                \\
    Number of periods                                             & 24                           \\
    Electrode spacing                                             & \SI{4}{cm}                   \\
    Electrode height                                              & \SI{20}{cm}                  \\
    Deflector shape                                               & cylindrical                  \\
    Radial bending $E$-field                                      & \SI{4.4}{MV/m}               \\
    Straight section length                                       & \SI{4.16}{m}                 \\
    Quadrupole length                                             & \SI{0.4}{m}                  \\
    Quadrupole strength                                           & \SI{\pm 0.21}{T/m}           \\
    Bending section length                                        & \SI{12.5}{m}                 \\
    Bending section circumference                                 & \SI{600}{m}                  \\
    Total circumference                                           & \SI{800}{m}               \\
    Cyclotron frequency                                           & \SI{224}{kHz}                \\
    Revolution time                                               & \SI{4.46}{\micro s}          \\
    $\beta_{x}^{\textnormal{max}},~ \beta_{y}^{\textnormal{max}}$ & \SI{64.54}{m}, \SI{77.39}{m} \\
    Dispersion, $D_{x}^{\textnormal{max}}$                        & \SI{33.81}{m}                \\
    Tunes, $Q_{x}, ~ Q_{y}$                                       & 2.699, 2.245                 \\
    Slip factor, $\frac{dt}{t}/\frac{dp}{p}$                 & -0.253                       \\
    Momentum acceptance, $(dp/p)$                                 & \num{5.2e-4}                 \\
    Horizontal acceptance [mm mrad]                               & 4.8                          \\
    RMS emittance [mm mrad], $\epsilon_{x}, ~\epsilon_{y}$        & 0.214, 0.250                 \\
    RMS momentum spread                                           & \num{1.177e-4}               \\
    Particles per bunch                                           & \num{1.17e8}                 \\
    RF voltage                                                    & \SI{1.89}{k V}               \\
    Harmonic number, $h$                                          & 80                           \\
    Synchrotron tune, $Q_{s}$                                     & \num{3.81e-3}                \\
    Bucket height, $\Delta p/p_{\textnormal{bucket}}$             & \num{3.77e-4}                \\
    Bucket length                                                 & \SI{10}{m}                   \\
    RMS bunch length, $\sigma_{s}$                                & \SI{0.994}{m}                \\
    \bottomrule
  \end{tabular}
  \label{tab:specs}
\end{table}%
\section{Systematic Error sources}\label{sec:systematicerrors}
The primary quantity of interest is the vertical spin precession rate
$dS_{y}/dt$ that lets us estimate the intrinsic dipole moment of the proton
$d_{p}$. The target sensitivity of $d_{p}={10^{-29}e\cdot\textrm{cm}}$
corresponds to a vertical spin precession rate of
$dS_{y}/dt = \SI{1}{n rad/s}$ (this number will be useful throughout the work).
Thus, any non-EDM originating vertical spin precession rate
larger than $\SI{1}{n rad/s}$ is considered a potential systematic error source.

Ideally, the EDM search is accomplished with positive helicity CR
100\% longitudinally polarized beams. Realistically, as little as $\approx 10^{-3}~\textrm{rad}$
average radial spin component would be uncontrollable due to statistical
limitations alone (see \Cref{sec:polarimeter}). Some systematics (e.g. Vertical
Velocity --- \Cref{sec:vertical_velocity}) are only sensitive to the radial spin
component --- $S_{x}$. Such systematics must always be considered not only due
to little average radial spin component being present (inadvertently), but also
due to free horizontal spin precession oscillations due to lattice imperfections.

The initial average spin direction, between maximally longitudinal and maximally
radial polarization directions, could be controlled --- the initial
$S_{s}/S_{x}$ ratio. Choosing this ratio is a powerful tool to clearly
differentiate the systematic error sources into longitudinal and radial
polarization originating types.

In the following subsections, relevant to the EDM search, potential systematic error
sources will be discussed. With the mentioned mixing of polarizations in mind,
the systematics pertaining to the radial polarization direction has its effect
reduced for the EDM search (longitudinal polarization) by at least a factor of $10^{3}$. The horizontal spin precession rate of the CR beams will be controlled using feedback with a combination of machine (RF-cavity) frequency and vertical magnetic trim fields.
A brief summary of the systematic error sources is given on \Cref{tab:systematics}.
\begin{table*}[tbp]
  \centering
  \caption{Summary of the main systematic error sources in storage ring EDM rings. ``T-BMT term'' indicates the
  driving term in \Cref{eq:omega_a} which lets us infer the sensitive
  polarization direction.} 
  \begin{tabular}[t]{p{0.15\linewidth} @{\hskip 0.5in} l @{\hskip 0.5in} p{0.6\linewidth}}
    \toprule
    Name & T-BMT term & Comments \\
    \midrule
    Radial magnetic field & 
    $ S_s \cdot B_x$ &
    Main systematic error source in the All-electric  ring design but not in rings with magnetic focusing.
    In rings with strong magnetic focusing (Hybrid (4-fold) and
    Symmetric-Hybrid) the external magnetic field is completely shielded out. Find
    more details in \cite{haciomeroglu_hybrid_2018}.\vspace{0.5cm}\\  

    Vertical electric field (dipole) & 
    $ S_s \cdot \beta_s \cdot E_y$ &
    Incorporation of CR beams completely eliminates this effect, as the vertical
    spin precession created by the vertical electric field is in opposite
    direction to the true EDM signal.  This is expected to be the largest systematic error source for rings without simultaneously stored CR beams. Trim vertical-electric-field plates, symmetrically distributed around the ring, will be used to keep the same sign vertical spin precession rate to zero for the CR beams.
    \vspace{0.5cm}\\ 

    Vertical velocity &
    $ S_x \cdot \beta_y \cdot E_x$ &
    Main systematic error source for DM/DE in Hybrid (4-fold) ring design and of a secondary concern for the EDM target sensitivity. By making the
    lattice symmetric --- Symmetric-Hybrid lattice (this work) --- this effect
    reduces by several orders of magnitude making it acceptable for the DM/DE target and completely negligible for the EDM sensitivity.  More discussion is found in
    \Cref{sec:vertical_velocity}.\vspace{0.5cm}\\ 

 Vertical electric field (quadrupole) & 
    $ S_s \cdot \beta_s \cdot E_y$ &
        All effects that depend on the CR beams separation, either in the vertical or in the horizontal direction, can be easily eliminated by artificially enlarging the separation with the application of small dipole magnetic fields at the magnetic quadrupole locations. By selectively splitting the CR beams using dipole correctors of the
    magnetic quadrupoles, the value of the parasitic electric (skew) quadrupole
    can be measured precisely and corrected. More details is found in
    \Cref{sec:spinbasedalignment}.\\
    
    \bottomrule
  \end{tabular}
  \label{tab:systematics}
\end{table*}

\subsection{Vertical Velocity}\label{sec:vertical_velocity}
The vertical velocity systematic error originates from the term proportional to
\begin{equation}
  \left(\vec{S} \times (\vec{\beta} \times \vec{E})_s\right)_y=  S_x \cdot \beta_y \cdot E_x
  \label{eq:vertical_velocity}
\end{equation}
in the spin dynamics equation --- \Cref{eq:dsdt,eq:dsdt2}. Non-zero radial spin component
$S_{x}$ (pointing inward/outward of the ring) combined with vertical velocity
$\beta_{y}$ may create vertical spin precession that would be indistinguishable
from EDM even with CR beam injection\footnote{Subscript $s$ indicates the
  direction along the ring azimuth.}.

Despite $\langle \beta_{y} \rangle \equiv 0$, the velocity would be non-zero if
averaged over the bending sections only ($E_{x}$ field regions). Formally, we
can only expect,
\[ L_{\textnormal{straight}}\langle\beta_{y}\rangle_{\textnormal{straight}} + L_{\textnormal{bending}}\langle\beta_{y}\rangle_{\textnormal{bending}} = 0, \]
each of the
$\langle\beta_{y}\rangle_{\textnormal{straight}}$ and $ \langle\beta_{y}\rangle_{\textnormal{bending}}$
might not be zero individually. This leads to a possibility of
\[ \langle \beta_y \cdot E_x \rangle \ne 0,\]
and,
\[ dS_{y}/dt \propto \langle S_{x}\cdot\beta_y \cdot E_x \rangle \ne 0,\]
which is the essence of the effect.

This systematic is also known as vertical orbit corrugation or the
``roller--coaster effect''. It is most prominent in the radial polarization case;
thus its effect is at least a few orders of magnitude less in longitudinal
(applicable to the EDM search) polarization. In order to isolate and understand this effect better, we
put the beam  in radial polarization and create vertical orbit corrugation by
vertically misaligning one magnetic quadrupole at a time.
A single vertically misaligned quad induces vertical imbalance that creates a non-zero
average vertical velocity.

\begin{figure}[tbp]
  \centering
  \includegraphics[width=0.99\linewidth]{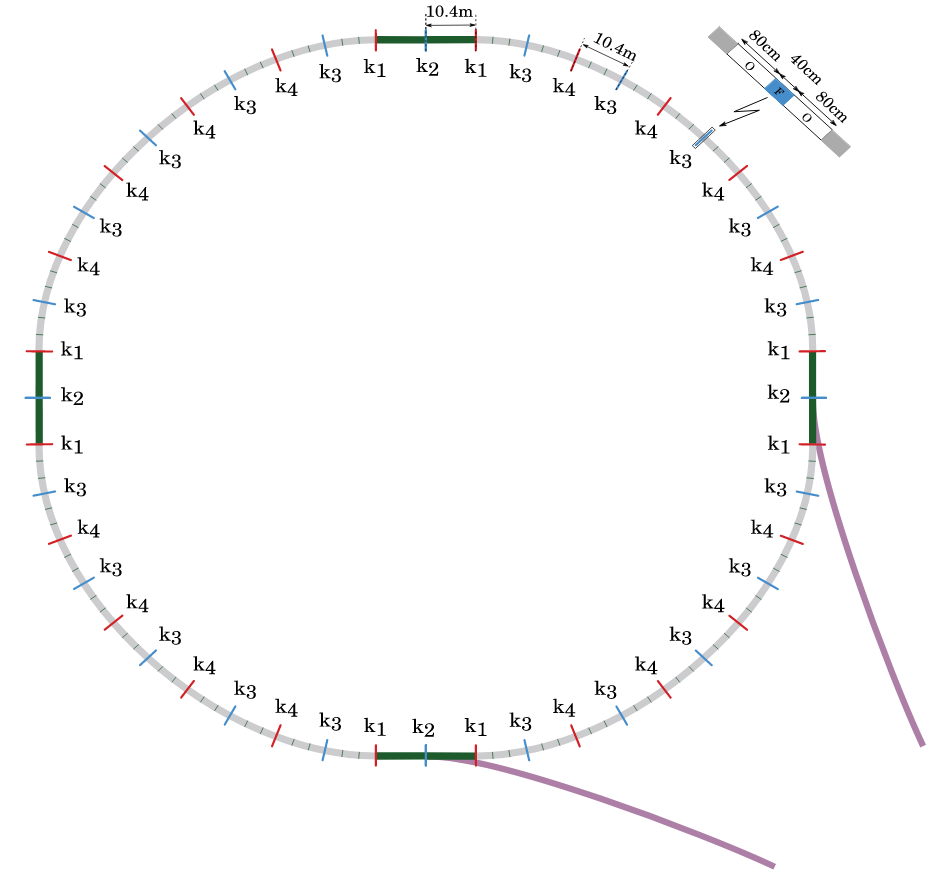}
  \caption{Hybrid (4-fold) ring design, the presence of the long straight sections severely
  reduce the number of symmetric points in azimuth (adapted from
  \cite{anastassopoulos_storage_2016}).}\label{fig:original-ring}
\end{figure}

\begin{figure}[tbp]
  \centering
  \includegraphics[width=0.99\linewidth]{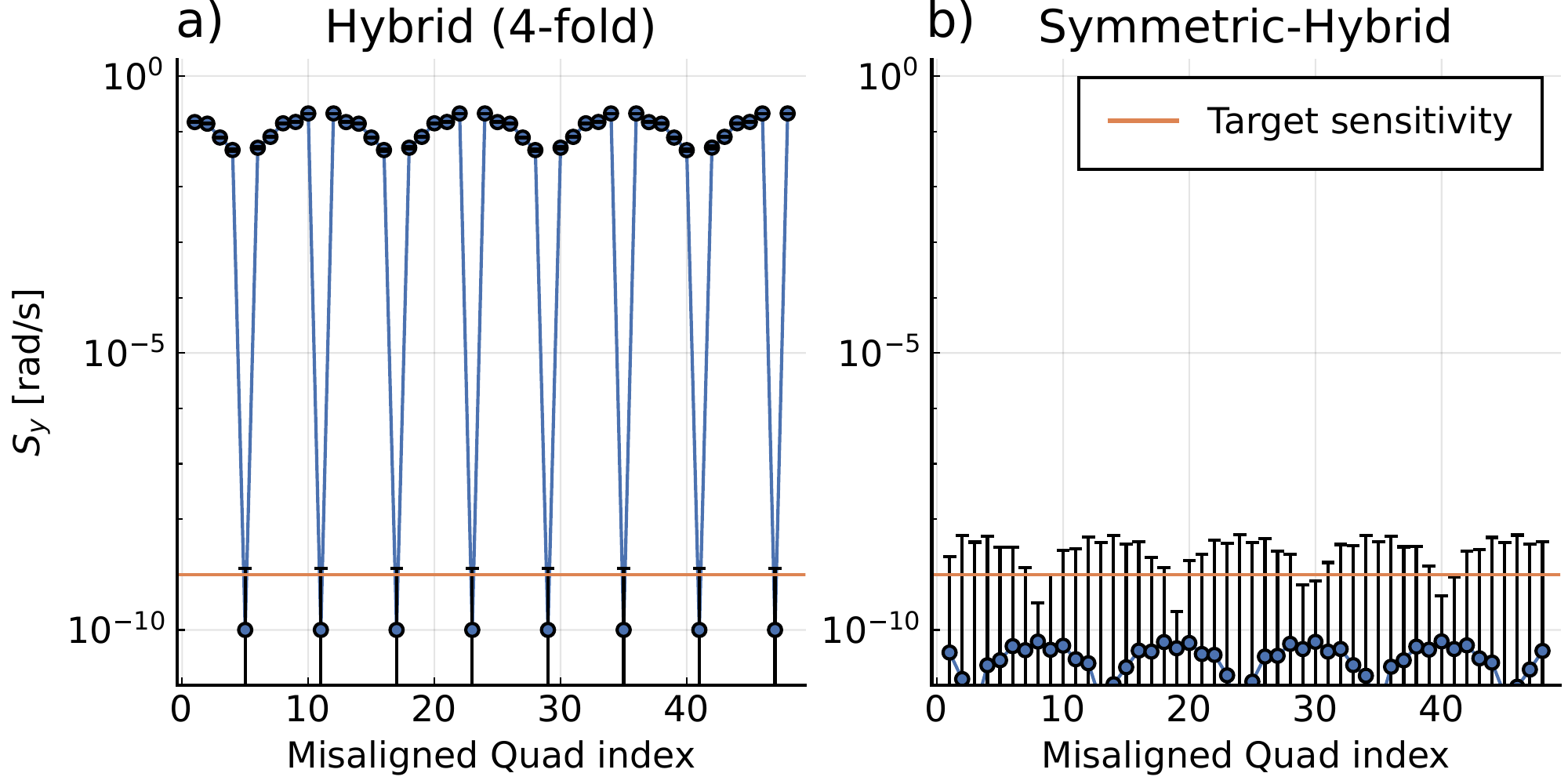}
  \caption{\emph{Radial polarization case $S_x=1$}, sensitive to DM/DE (even though the vertical velocity effect directly only affects the DM/DE sensitivity, if it is too large it will inevitably affect the EDM sensitivity of the stored bunches with primarily longitudinal polarization as well).  Vertical spin precession rate vs.\
    index of the $ \SI{100}{\micro m}$ vertically misaligned quad (one at a time) along the azimuth. The
    orange lines correspond to the target EDM sensitivity.
    \\(a) The original Hybrid (4-fold) ring design is used
    (\Cref{fig:original-ring}). Dips of the
    graph correspond to quads in the center of the four long straight sections 
    (in green color) shown in \Cref{fig:original-ring} and the quads maximally away from 
    long straight sections.
    \\(b) The Symmetric-Hybrid ring design is used (\Cref{fig:ccw-cw}). Notably the
    performance is many orders of magnitude better than the Hybrid (4-fold) ring (a). Simulations with lattice parameters slightly off their ideal values do not seem to show a significant deterioration of  the cancellation factor. 
    \\ The bottom part of some error bars go beyond zero; thus are invisible in log-scale plot. 
    Such large error bars hint that the true underlying precession rate is zero with large oscillations. 
    Error bars arise due to an inability to determine the exact vertical precession rate from finite digital
    data (numerical tracking). There is more about this in 
    \Cref{sec:data}. 
    }\label{fig:quad-misalignment-both}
\end{figure}

The vertical velocity systematic is especially prominent in ring designs where all the
quadrupoles are not equivalent in misalignments with respect to each other. For
example, the Hybrid (4-fold) ring design
\cite{haciomeroglu_hybrid_2018}, (\Cref{fig:original-ring}), where misaligned
quads are not equivalent (symmetric), shows clear islands of tolerance to
vertical quad misalignments, \Cref{fig:quad-misalignment-both} (a). Only the
quads at locations where the ring looks the same in both directions
longitudinally, the Hybrid (4-fold) ring is insensitive to the corresponding
misalignments (dips on  \Cref{fig:quad-misalignment-both} (a)).

By making the ring symmetric for all the quads longitudinally (\Cref{fig:ccw-cw}),
all the quadrupoles are made equivalent and thus
tolerant to vertical misalignments --- \Cref{fig:quad-misalignment-both} (b).
The quadrupoles were misaligned one at a time by $\SI{100}{\micro m}$ which
splits the CR beams by around $\SI{250}{\micro m}$. Vast reduction of the 
background vertical precession rate is achieved
(\Cref{fig:quad-misalignment-both} (b)) with the Symmetric-Hybrid ring design,
therefore reducing the systematic error source by a few orders of magnitude.

Distributing the quadrupole misalignments randomly with rms $\sigma=\SI{100}{\micro m}$ 
that leads more than $\SI{1}{mm}$ CR beam separation has been tested too. Background
vertical spin precession with radially polarized beam does not exceed
$\SI{1}{nrad/s}$ in the Symmetric-Hybrid lattice. 


\subsection{Dipole $E$-field}\label{sec:dipole_e}
The dipole $E$-field systematic originates from the
\begin{equation}
  \qty(\vec{S}\times\qty(\vec{\beta}\times\vec{E})_{x})_{y}=S_{s}\cdot\beta_{s} \cdot E_{y}
  \nonumber
\end{equation}
term in \Cref{eq:dsdt,eq:dsdt2}. A non-zero $E_{y}$ could arise due to some tilt ($x-y$
plane rotation) in the deflector plates. Each bending section, being randomly
tilted, contributes to the average non-zero dipole $E$-field initially
present in the storage ring.
$E_{y}$ creates an EDM-like signal for one
of the counter rotating beams. However, the true EDM-signal causes a
vertical spin precession in opposite directions for CR beams. The difference of precession rates for CR beams gives
us the true EDM signal, as the dipole $E$-field creates a discernible from EDM
signal with CR beam storage --- \Cref{fig:e_multipoles} (N=0). Formally,
\[
  \qty(\frac{dS_{y}}{dt})_{\textnormal{EDM}} = \frac{1}{2}\qty(\frac{dS_{y}}{dt})_{\textnormal{CW}} - \frac{1}{2}\qty(\frac{dS_{y}}{dt})_{\textnormal{CCW}}.
\]
More about spin data combinations is given in \Cref{sec:data_combination}.

An average background $E_y$ --- \Cref{fig:e_multipoles} ($N=0$) --- creates a large
spin precession in both CR beams. Such a large spin precession, but the same for CW and CCW beams,
 is undesirable for a multitude of reasons. In practice, a trimming
$E_y^{\textnormal{trim}}$ dipole electric field will be applied to compensate
for the large spin precession such that no discernible spin precession
($<\SI{e-6}{rad/s}$) is seen in both CR beams that effectively sets
$E_y+E_y^{\textnormal{trim}}=0$. Gradual adjustment of
$E_y^{\textnormal{trim}}$ will eliminate same direction (non-EDM like) vertical
spin precession in both CR beams.
 

\subsection{Quadrupole $E$-field and Spin-based alignment}\label{sec:spinbasedalignment}
\begin{figure}[tbp]
  \centering
  \includegraphics[width=0.99\linewidth]{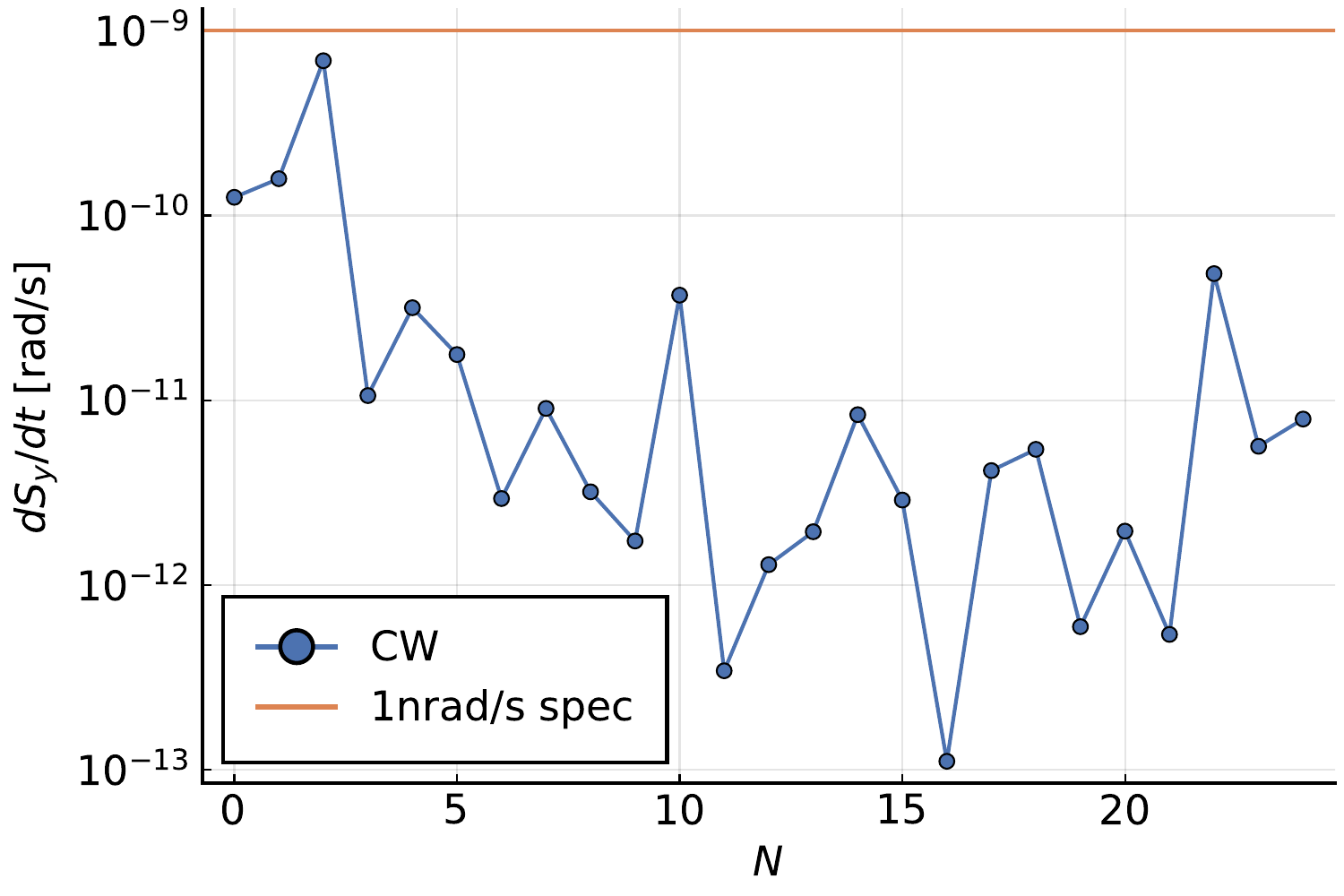}
  \caption{\emph{Longitudinal polarization case $S_s=1$, CW beam only.}  Vertical spin
  precession rate vs $B_x=\SI{1}{\nano \tesla}$ field $N$ harmonic around the
  ring azimuth. The magnetic field amplitude is chosen to be similar to beam
  separation requirements in \Cref{sec:realistic}, more than $B_x=\SI{1}{\nano
  \tesla}$ splits the CR beams too much.\\ \upperlimit}\label{fig:b_multipoles}
\end{figure}

In the absence of vertical electric fields $E_{y}^{\textnormal{external}}=0$,
any non-zero $B_{x}^{\textnormal{external}}$ would be compensated by a magnetic
force coming from quadrupoles; therefore, it would on average result in
$\langle B_{x}\rangle=0$. Magnetic fields are balanced
by magnetic fields; hence, there is no apparent vertical spin precession
due to $B_{x}^{\textnormal{external}}$ for all the $N=1,2,3,...,24$ harmonics
--- \Cref{fig:b_multipoles}. 

However, in case $E_{y}^{\textnormal{external}}\ne 0$, $\langle B_{x}\rangle = 0$
is no longer guaranteed. We can only expect to  first order,
omitting ``external'' superscript\footnote{$\beta_{s}$ can safely be assumed
constant as its variation is negligibly small.}, 
\begin{equation}
  F_{y}  = q \qty(E_{y} + c\beta_{s}B_{x}) = 0.
  \label{eq:balance}
\end{equation}

\Cref{eq:balance} needs to be true on average for the closed orbit. But, zero
on average does not guarantee local absence of electric and magnetic forces. In
order to prevent parasitic vertical spin precession due to
$E^{\textnormal{quad}}_y$ and $B_x$, all the multipoles and harmonics need to be
addressed individually.

\begin{figure}[tbp]
  \centering
  \includegraphics[width=0.99\linewidth]{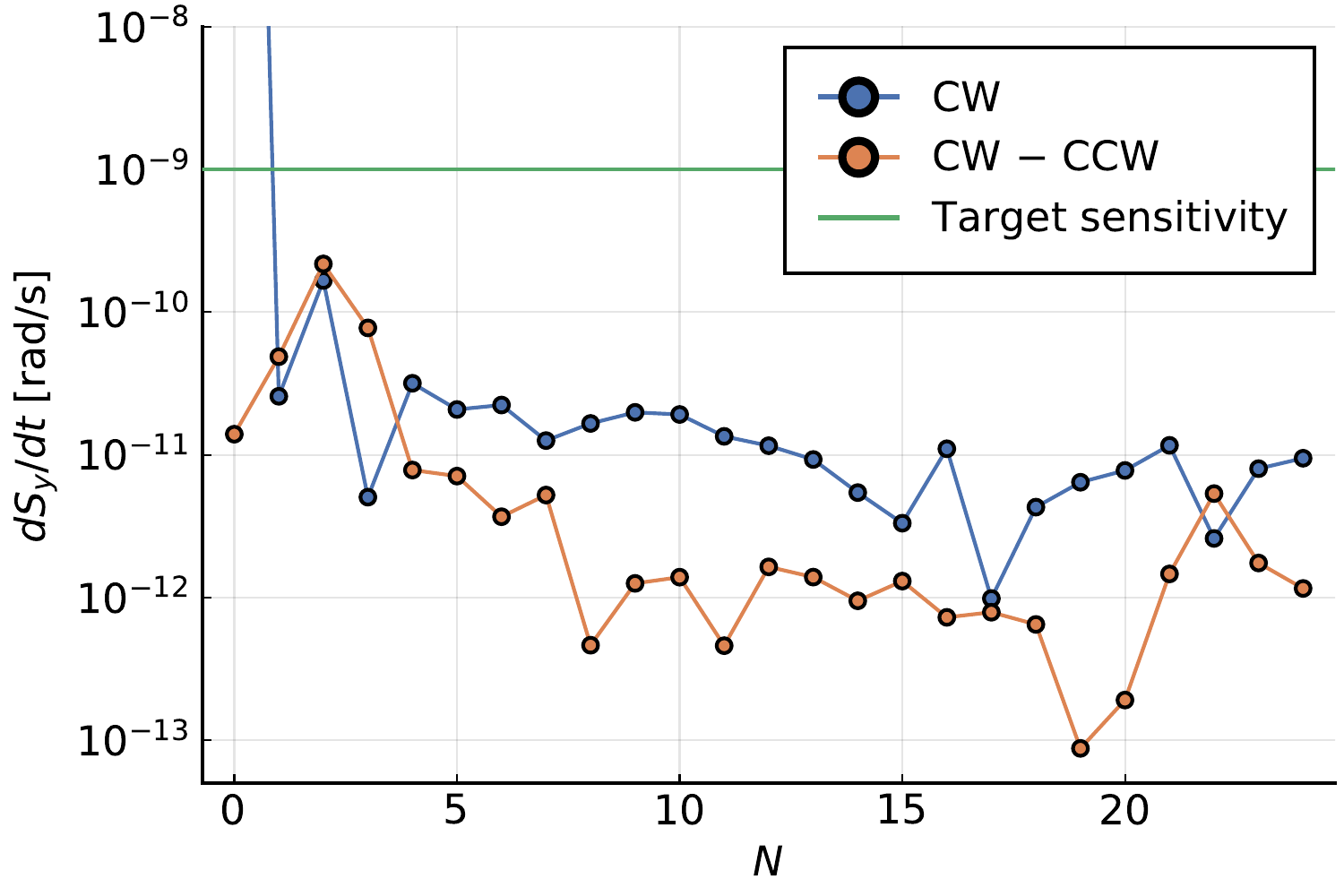}
  \caption{\emph{Longitudinal polarization case $S_s=1$}, sensitive to EDM. Vertical spin
    precession rate vs $E_y=\SI{10}{V/m}$ field $N$ harmonic around the ring
    azimuth. For $N=0$, the precession rate for CW (or CCW) beam is around
    $\SI{5}{rad/s}$. The difference of the precession rates for CR beams
    (orange) is below the target sensitivity for all $N$.
    \upperlimit}\label{fig:e_multipoles}
\end{figure}

The most dominant multipole, the dipole ($E^{\textnormal{}}_{y}$)
and all its harmonics do not create EDM-like signal due to the simultaneous CR beam
storage (\Cref{fig:e_multipoles}). This is also true for higher odd multipoles ---
i.e. sextupole, decapole, 14-pole, etc.

The next multipole -- the quadrupole ($E^{\textnormal{quad}}_y$) and higher-order
even multiples, i.e. octupole, 12-pole, etc., need to be addressed separately.
For example, if CR beams are separated on average by $\pm
\Delta y$ due to external $B_x$ field and some parasitic quadrupole
$E^{\textnormal{quad}}_y(y)=K_{e}y$ is present, the vertically separated beams
would experience electric field in opposite directions $E^{\textnormal{quad}}_y=\pm
K_e \Delta y$; therefore, an EDM-like vertical spin precession is observed.

Whenever an electric field balances a magnetic field and vice
versa, a vertical spin precession might take place. The All-electric ring design is
completely immune to stray electric fields, but highly sensitive to radial magnetic
fields. The Hybrid (4-fold) and Symmetric-Hybrid designs, in contrast, are sensitive
to electric fields. However, the effect of the main multipole --- dipole
$E_y$ field --- is distinguishable from the true EDM signal with CR beam storage.

The presence of $E^{\textnormal{quad}}$ can be monitored by controlling
$B_{x}$\footnote{more about measuring $B_{x}$ in \Cref{sec:squid}.}. The
combination of $E^{\textnormal{quad}}$ and $B_{x}$ produces non-zero vertical
spin precession rate ${dS_{y}}/{dt}$. $B_{x}$ could be made large on purpose,
for example by controlling dipole correctors of the magnetic quadrupoles. Being
able to freely control $B_x$ and all of its harmonics\footnote{Due to the low
tune $(\approx 2)$ only a few harmonics need to be probed.} lets us selectively
(for each $N$ harmonic) amplify and then reduce the effect of initially unknown
$E^{\textnormal{quad}}$.


Likewise, skew electric quadrupoles $E^{\textnormal{skew q.}}_y(x)=K_{e'}x$
couple to vertical magnetic field $B_y$ to create false EDM signals. Measurement
of $E^{\textnormal{skew q.}}$ is performed with the same procedure. By purposely
introducing $B_y$ using the dipole correctors of the magnetic quadrupoles,
measurement of  ${dS_{y}}/{dt} \propto E^{\textnormal{skew q.}} \times B_y$ lets
us infer the value of $E^{\textnormal{skew q.}}$ for all relevant $N$ values.

Similarly, image charge, beam-beam, etc.~effects that may produce quadrupole or
higher electric field multipoles are treated the same way. We do this since the effect on the
vertical spin precession rate does not depend on the origin of the electric
field.

The presented idea of controlling the $E$ fields using spin measurements, which
are extremely sensitive, is labeled as ``Spin-based alignment'' (SBA).  Leveling
the ring to a high order using SBA is performed using various combinations of
the bunch polarizations. For example, radial polarization bunches, being
sensitive to Vertical Velocity (\Cref{sec:vertical_velocity}), will be used as a
feedback to measure the vertical orbit corrugation. Other spin polarization
directions such as vertical polarization can be used to test the effects of
Geometrical phase and other as yet unknown systematics.

In principle, SBA could be used in other accelerator facilities that
require precise ring alignment. The Spin dynamics is much more sensitive to
EM-fields than the beam dynamics; thus, it can serve as a sensitive probe of 
lattice imperfections.

\subsection{Geometrical Phase}
Geometrical (Berry) Phase effect, as it is known in its most common definition,
\cite{berry,PhysRevLett.97.131801,PhysRevA.70.032102} is attributed to an extra
acquired phase difference when a given system undergoes a cyclic adiabatic
process.

In the context of storage ring EDM experiments unwanted spin precession, obtained
due to non-commutativity of  successive rotations, is referred to as Geometrical
Phase. The spin precession is proportional to the product of successive rotation
amplitudes.

\begin{figure*}[tbp]
  \centering
  \includegraphics[width=0.99\linewidth]{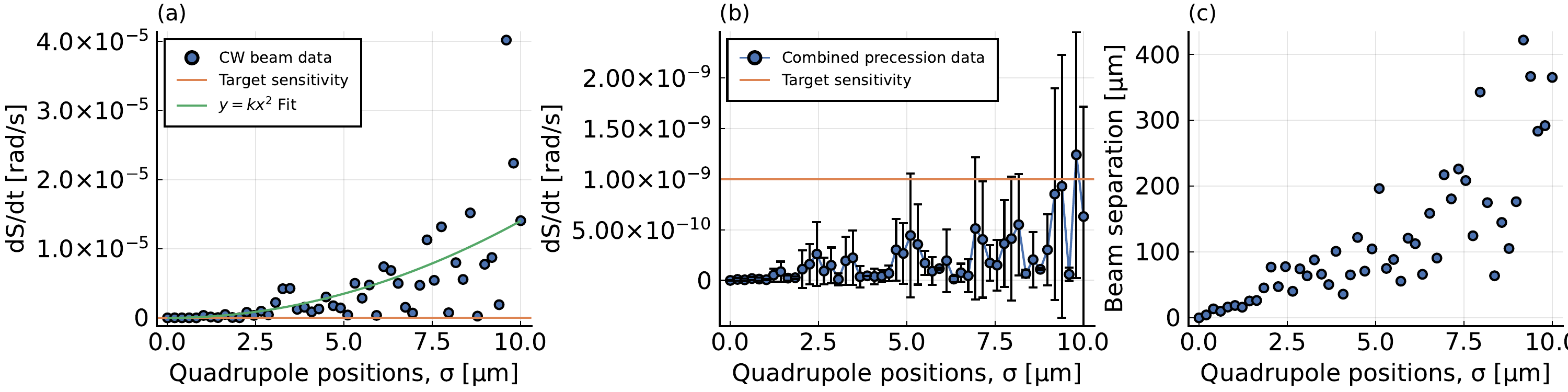}
  \caption{(a) \emph{Longitudinal polarization case, CW beam only.} Vertical spin precession rate (absolute) 
  vs.~random misalignments of quadrupoles in both $x,y$ directions by rms $\sigma$ with 
  different seeds per each point (when the same seeds are used everywhere the $y=kx^2$ fit is perfect, meaning that every point can be extrapolated to any rms $\sigma$ value using this functional form). Combination with CCW and quadrupole polarity switching
  achieves large cancellation --- see (b).
  (b) \emph{CW and CCW beam and with quadrupole polarity switching.} Total combination as 
  presented in \Cref{sec:data_combination}. Notably, the background vertical spin precession
  rate (absolute) stays below the target sensitivity. Irregularity of the points is discussed in \Cref{sec:data}. 
  (c) Correspondence between CR beam separation and rms $\sigma$ quadrupole misalignments.
  }\label{fig:mag-geometrical}
\end{figure*}

The product dependence is verified by linearly increasing the amplitude of
successive rotations in the $x, y$ plane. This is accomplished by misaligning
all magnetic quadrupoles randomly with rms $\sigma$ (both $x,y$ directions). By
increasing $\sigma$ while observing the growth of the unwanted vertical
precession rate, the square dependence is favored ---
\Cref{fig:mag-geometrical} (a). Significant cancellation is achieved by
incorporating both CR beams \cite{silenko_berry_2018} including runs with reversed magnetic
quadrupole polarities --- \Cref{fig:mag-geometrical} (b). As it is apparent from --- \Cref{fig:mag-geometrical} (c), even small quadrupole magnet misalignment causes large CR beam splitting, which will be finely reduced to well below $100$~\si{\micro m} by applying dipole correction B-fields at the quad locations.

Since the quads are misaligned randomly, it is not immediately clear what is causing the total effect.
A thorough study using a straight lattice\footnote{A straight lattice stripped of electrostatic bends, consisting of 
only quadrupoles. It is essentially of infinite length with 800m periodicity.} was performed to reveal that 
random misalignments of quads alone do not cause vertical spin build-up. Hence, we conclude that the 
vertical spin precession in one direction (CW --- \Cref{fig:mag-geometrical} (a)), arises due to intermixing with
other systematic effects such as vertical velocity, and other second order systematics (some are discussed in \Cref{sec:geom_phase_add}).

Numerical tracking shows that the EDM-like vertical spin
precession caused due to Geometrical Phase is insignificant  when the
CR beam separation is below a few hundred \si{\micro m} (corresponding to quadrupole misalignments of around 
few \si{\micro m}) --- \Cref{fig:mag-geometrical} (b, c),
whilst orbit planarity even around $\SI{10}{\micro m}$ were
achieved~\cite{yang_high-resolution_2006,shiltsev_space-time_nodate}  by
mechanical means using water levels. More about beam separation measurements appears in \Cref{sec:squid}.

\subsection{General Relativity}
General relativistic (GR) effects caused by  gravity
and rotation of the Earth can be observed in high-precision
experiments. The spin dynamics in the considered pEDM experiment could be
affected. In connection with the Equivalence Principle, one can
always introduce a \emph{local} Lorentz (anholonomic) coordinate system based on
a tetrad of appropriate orthogonal coordinate vectors. Dynamics of the momentum
and spin in this coordinate system is defined by equations of motion formally being the same
with the usual equations given by electrodynamics in the Minkowski
spacetime~\cite{PRDThomas,PRD2016,PRD2017}. The general relativistic effects in
storage ring EDM experiments have been analyzed in previous
studies~\cite{PRD2017,PRD2007,OFS,LaszloZimboras,NikVerg} and the corresponding
systematic corrections have been calculated. It has been explicitly shown in
Ref.~\cite{NikVerg} that the final results obtained in these studies perfectly
agree with each other. The net effect due to GR creates a little or
a distinguishable-from-the-EDM signal with CR beam storage. Hence, GR related effects
are not significant for the current proposal.

\subsection{Spin coherence time}\label{sec:sct}
The spin coherence time (SCT), also known as in plane polarization (IPP)
lifetime, is essential to achieve the desired
sensitivity requirements~\cite{edm_proposal}. An EDM search with a longitudinally
polarized beam requires a SCT of around $\SI{e3}{s}$. It has been shown that long
SCTs are correlated with zero chromaticity
conditions~\cite{vasserman_comparison_1987,koop1988spin}. Chromaticity is
defined as,
\begin{equation*}
  \xi_{x,y} = \frac{\delta Q_{x,y}/Q_{x,y}}{\delta p/p}.
\end{equation*}

Sextupoles can reduce chromaticity, and hence make it possible
to achieve a long $\approx \SI{e3}{s}$ SCT, as it has been experimentally shown at
COSY~\cite{jedi_collaboration_how_2016}. However, our studies show that in a ring with electric bending, as opposed to rings with regular magnetic bending, the required conditions are not exactly the same. Therefore, we focused directly searching for a long SCT.

A long SCT could be achieved by using magnetic sextupoles. Magnetic sextupole
fields are defined as,
\begin{align*}
    B_x &= 2 k^{m} x y \\
    B_y &= k^{m} ( x^2 - y^2 ).
\end{align*}
A magnetic sextupole pair $k^{m}_{1}, k^{m}_{2}$ overlaps with the magnetic
quadrupoles (\Cref{fig:FODO}) in order not to break the symmetry requirements of
the lattice. The reference particle horizontal precession rate is significantly
improved if the correct sextupole fields are used --- \Cref{fig:k1_k2}. 

The sextupoles could also be electric instead,
\begin{align*}
    E_y &= -2 k^{e} x y \\
    E_x &= k^{e} ( x^2 - y^2 ).
\end{align*}
A similar spin precession behavior is seen--- \Cref{fig:elec_k1_k2}.

However, the optimal pair (electric or magnetic) of sextupole strengths
$k^{m,e}_{1}, k^{m,e}_{2}$ (\Cref{fig:k1_k2,fig:elec_k1_k2}) is not symmetric
with respect to CR beams. It first seems that the SCT can be improved for one beam direction
only. It is an acceptable solution if we allow the beams to be injected
separately while adjusting the sextupole strengths accordingly for each of the
beam directions.

Alternatively, by incorporating magnetic and electric
sextupoles at the same time, the SCT could be improved for both CR beams. The symmetry of the problem
(\Cref{fig:k1_k2,fig:elec_k1_k2}) shows that CR beams experience the
same effect from magnetic and electric sextupoles in case,
\begin{alignat}{2}
  k^m&= k^m_1 & &=-k^m_2 \nonumber \\
  k^e&= k^e_1 & &=k^e_2. \label{eq:ke_km}
\end{alignat}
Hence, having both magnetic and electric sextupoles that follow \Cref{eq:ke_km}
will lead to a better SCT for both CR beams --- \Cref{fig:hybrid_k1_k2}.

Although the optimal sextupole pair found in
\Cref{fig:k1_k2,fig:elec_k1_k2,fig:hybrid_k1_k2} is for the reference particle,
the same pair happens to be near optimal for the corresponding bunch of
particles too.

\begin{figure}[tbp]
  \centering
  \includegraphics[width=0.99\linewidth]{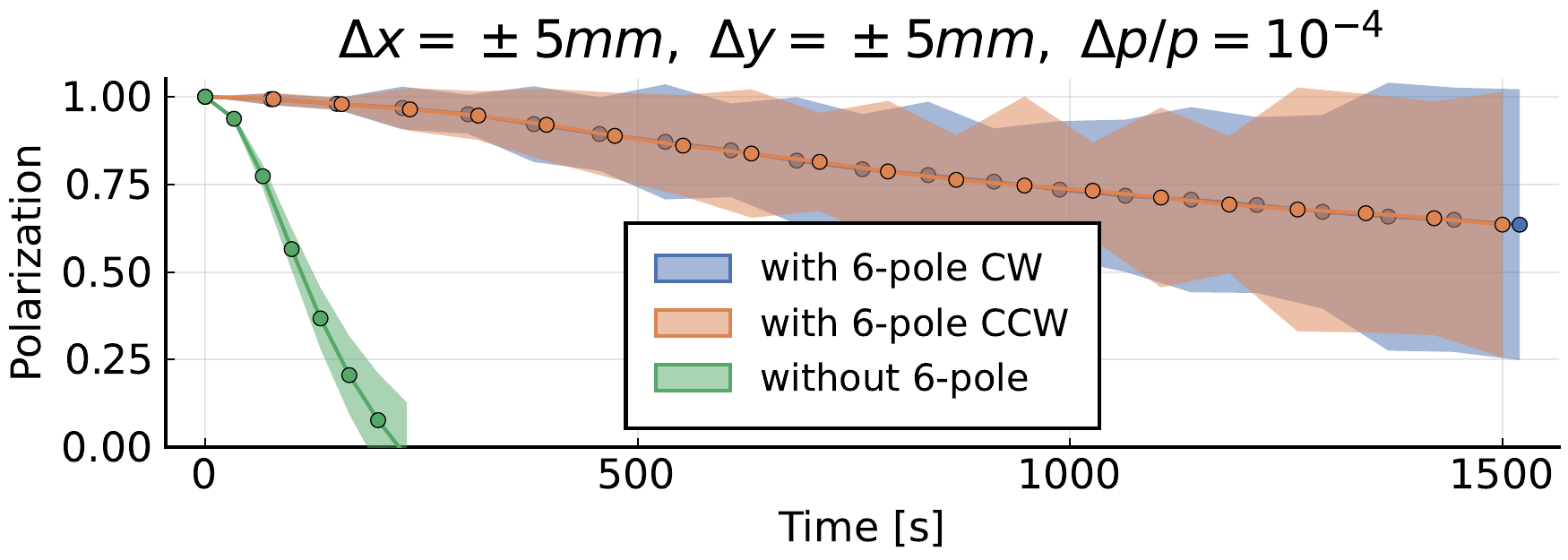}
  \caption{Magnitude of the polarization vector vs. time simulation with a
    realistic bunch structure given as, $\Delta p/p = 10^{-4}$ rms;
    $\Delta x/x=\pm \SI{5}{mm}, \Delta y/y = \pm \SI{5}{mm}$ as maximum and assuming a uniform phase space distribution. The polarization
    retains a high value with hybrid (magnetic and electric) sextupoles for both
    CW (blue) and CCW (orange) bunches compared to the nominal case without
    sextupoles (green). 
    The estimated SCT is expected to become longer when the betatron amplitudes and 
    momentum exchanges in three dimensions, due to IBS, are taken into account.
    The simulation is subsecond long, with the polarization at
  $t \gg \SI{1}{s}$ estimated by measuring the precession rate for each particle in the simulation, 
  then extrapolated with the corresponding error propagation. Vertical ribbon bands 
  indicate the digitization uncertainty as discussed in \Cref{sec:data}.}\label{fig:bunch_sct}
\end{figure}

By incorporating the best pair $k^{m}\qq{and} k^{e}$, the SCT improves vastly
for both of the CR beams at the same time for a bunch of particles ---
\Cref{fig:bunch_sct}.  Additional details about finding the optimum sextupole
strengths are given in
\Cref{sec:optimum_sextupole}.

\subsection{Polarimeter systematic issues}\label{sec:polarimeter}
\begin{figure}[tbp]
  \centering
  \includegraphics[width=0.99\linewidth]{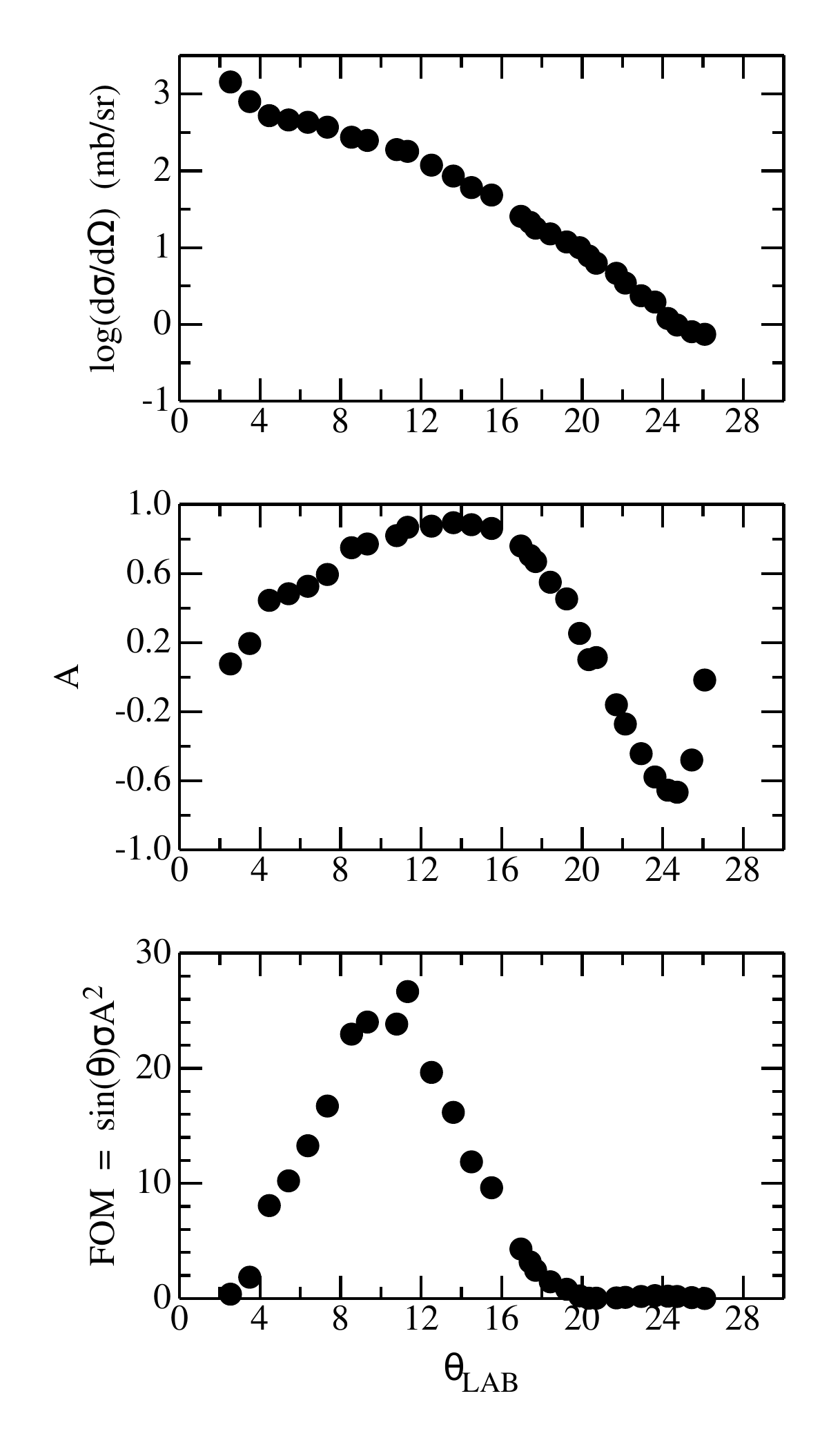}
  \caption{Angular distributions for elastic proton-carbon scattering at 250 MeV
    \cite{meyer1988proton} showing the differential cross section, analyzing power, and figure of
    merit. The figure of merit (FOM) indicates the statistical significance of utilizing parts of the angular distribution in a polarimeter; polar angles between $4^{\circ}$ and $16^{\circ}$ are optimal.
}\label{fig:polarimeter}
\end{figure}

Measurement of the proton beam polarization will most likely involve the
observation of the asymmetry in the elastic scattering of protons from a
light-mass target such as carbon. The differential cross section for protons is
given by
\[\sigma(\theta)_{\textnormal{Pol}}=\sigma(\theta)_{\textnormal{UNP}}\qty[1+pA(\theta)\sin{\beta}\cos{\phi}]\]
where $\theta$ is the polar scattering angle for the detected protons, and
$\beta$ and $\phi$ are the
polar and azimuthal angles for the proton polarization direction ($\phi$ measured from
the perpendicular to the scattering plane). $A$ is the analyzing power, which
describes the degree of sensitivity of the scattering to polarization acting
through the spin-orbit interaction between the proton and the nucleus. $p$ is the
beam polarization. At the energies expected for the EDM search, the small-angle
cross section and analyzing power are both large (\Cref{fig:polarimeter}).

The EDM signal arises from beam polarizations $(p)$ that are perpendicular to the ring plane. These may be detected by comparing the elastic scattering rates on opposite sides of the beam in the ring plane. The $\cos{\phi}$ azimuthal dependence produces opposite scattering rate changes in these two detectors. If the scattering rates are designated by L and R for the two sides of the beam and measurements are made with both $+$ and $-$ states of the beam polarization, then the vertical component may be determined from
\[
  \bar{p} = \frac{1}{A}\frac{r-1}{r+1} ~~~ r^{2} = \frac{L_{+}R_{-}}{L_{-}R_{+}}
\]
The combination of simultaneous left- and right-side detection with data using
opposite polarization states cancels many first-order errors in this analysis.

Accurately measuring small polarization rotations at the level of
$\si{\micro rad}$ means
being able to handle errors beyond the first order. To do this, we must create a
model of the terms driving these errors in order to provide a means of making corrections
for them in real time if possible. Such a model was created for the original
polarimeter used in beam studies at COSY~\cite{polarimeter2}. There must also be
parameters that scale the corrections that are themselves sensitive in first
order to the driving terms. One such choice is,
\[
  \phi = \frac{s-1}{s+1} ~~~ s^{2} = \frac{L_{+}L_{-}}{R_{+}R_{-}}
\]
which is sensitive to geometric errors in first order but not to the
polarization and
\[
  W = \frac{dL_{+}}{dt} +\frac{dR_{+}}{dt} + \frac{dR_{-}}{dt} + \frac{dL_{-}}{dt}
\]
which is sensitive to the sum of the detector count rates for correcting
rate-dependent errors. Next a calibration must be performed of the sensitivity
of the polarimeter to various orders of angle/position errors as a function of
these two driving terms. Once in place, monitoring the magnitude of these two terms
allows a correction to be made to any polarization observable in real time. This
was tested at COSY and proved correct to a level of $10^{-5}$ (limited by statistics)
with no suggestion that the method was encountering a limit.

There are a number of systematic effects that rely on the comparison of
asymmetries measured with CR beams. Most likely, this
will mean two sets of forward detectors mounted on either side of a single
target that is shared by the two beams. For elastic scattering from carbon,
backscattering from the target is usually less than $10^{-7}$ of the forward
scattering rate and should not be an issue. But the two polarimeters will be
separate instruments, and the calibration of their response to polarization must
be precise enough that the difference of the asymmetries they yield is
meaningful at the level of $10^{-6}$, what is needed for the EDM search.

\section{Discussion and Conclusion}\label{sec:discussion}
\subsection{Simulation with realistic conditions}\label{sec:realistic}
\begin{figure}[tbp]
  \centering
  \includegraphics[width=0.99\linewidth]{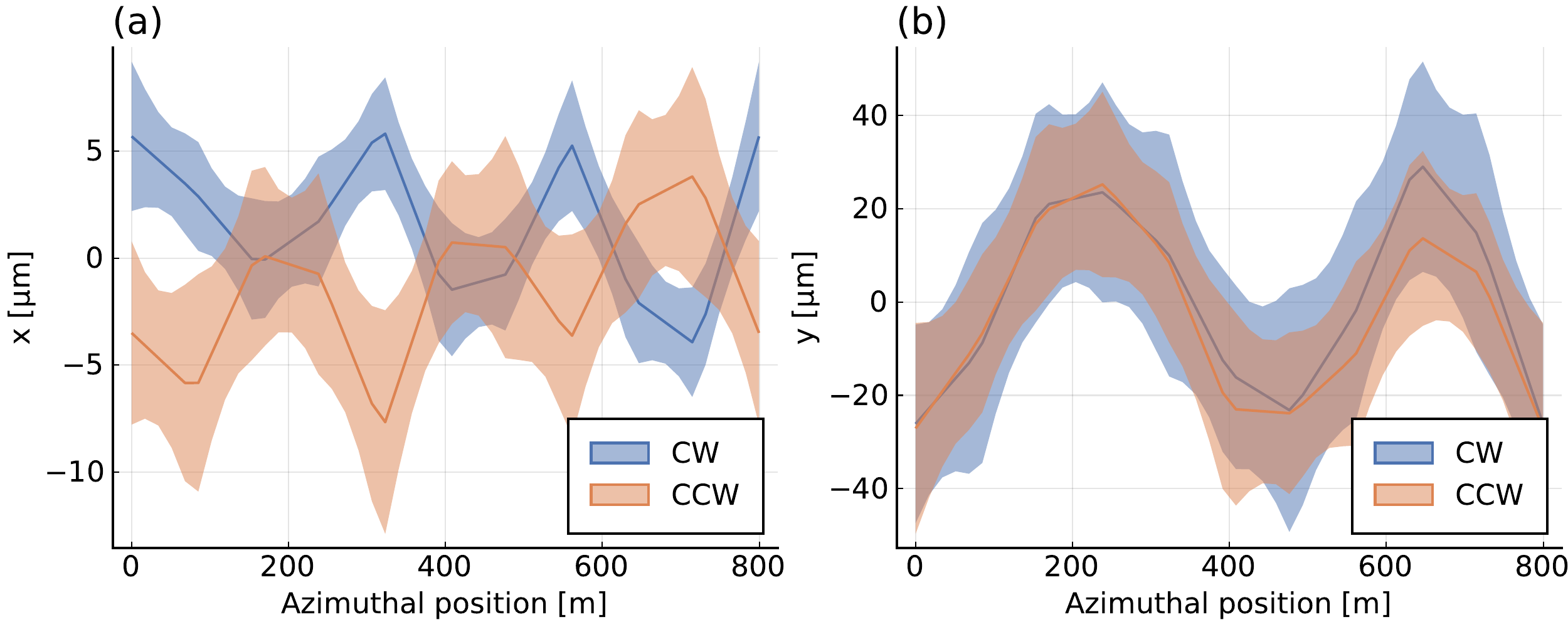}
  \caption{Single particle position averaged over \num{5e5} turns  split onto 48 bins. (a) Horizontal position throughout the ring azimuth. (b)
    Vertical position throughout the ring azimuth. Fill color shows
    standard deviations at the bins, roughly giving an idea about the spread.}\label{fig:azimuthal}
\end{figure}

We further demonstrate the feasibility of the experiment by including multiple
lattice imperfections such as both horizontal and vertical quadrupole
misalignments and deflector tilts. All in all, CR beams are required to
vertically overlap within $\pm\SI{5}{\micro m}$\footnote{A much larger beam separation can be tolerated, but $\SI{10}{\micro m}$ should be possible to achieve based on technology similar to the SQUID-based beam-position-monitors (S-BPMs) with a resolution of 10~nm$/\sqrt{\rm Hz}$~\cite{SBPM_1}.} with $\pm\SI{50}{\micro m}$
overall vertical closed orbit planarity. With such conditions, we first
numerically verify that the established realistic conditions are met --
\Cref{fig:azimuthal}. Then, vertical magnetic field $B_{y}$ and RF cavity
frequency is adjusted until no discernible (less than
$\sim\SI{1}{\micro rad/s}$) horizontal spin precession is present. Lastly,
we run with both normal and reversed magnetic quadrupole polarities and look at
the total EDM signal which is calculated as in \Cref{eq:edm_signal} below.

Upon examining the result -- \Cref{fig:total_precession}, it is clear that the
unwanted background residual EDM-like signal is below the target experimental
sensitivity; hence, the systematic error sources with such
lattice alignment requirements are low enough to allow the measurement of the proton
EDM to $d<10^{-29}e\cdot \mathrm{cm}$.

\subsection{$B$-field measurement}\label{sec:squid}
Although the Symmetric-Hybrid and Hybrid (4-fold) lattice designs completely
shield the beam from external magnetic fields, some limits to $B$-fields are
necessary due to the maximum beam splitting requirements. This section discusses the
technique of measuring the beam splitting which is also equivalent to measuring the magnetic fields experienced by the CR beams. 

With a specification that the CR beams can
split maximum up to $\pm \SI{5}{\micro \metre}$ ($\SI{10}{\micro \metre}$ in
total), only around sub-nT level B-field can be tolerated. This can be achieved
by a variety of techniques one of which is described here. The ring will be surrounded by sets of fluxgate
magnetometers and Helmholtz coils to eliminate the external field by
active cancellation. The number of sets located around the ring determines the azimuthal B-field harmonics
that can be probed and cancelled. The magnetic focusing system, if perfectly aligned, does not cause any splitting between the CR beams. Small external magnetic fields are also shielded by the focusing system, as shown in \Cref{fig:b_multipoles}. Since a typical quad field gradient is about 0.2~T/m, even a small quad misalignment will cause a large beam splitting and is expected to be the dominant source of the beam separation around the ring.

The
split can be measured by means of magnetic pick-ups. A $ \si{\micro \metre}$
level vertical split induces roughly $\si{\pico \tesla}$ level radial magnetic
field at a few $\si{cm}$ distance from the beam~\cite{SBPM_1}, due to the CR beams.
In order to increase the SNR, the quadrupole fields will be modulated at
$1-10~\si{kHz}$ by 1\%, which is coined as K-modulation~\cite{k_mod}. The
measurement can be easily accomplished with commercially available fluxgate magnetometers
(with a few pT/$\sqrt{\si{Hz}}$ sensitivity) operating at room temperature, while there are a variety of other
commercial options as well. A recently developed SQUID-based BPM has a potential
to measure the split with better than $\SI{10}{\femto\tesla}/\sqrt{\si{Hz}}$
sensitivity~\cite{SBPM_1}.

\subsection{Experimental knobs}
In this section, a brief summary of the available experimental knobs that
reduce the effects of systematic error sources is listed. Methods unused in this study
are marked and will require additional detailed studies.
\begin{itemize}
  \item \emph{CR beam storage}. Simultaneous CR beam storage eliminates a whole
        class of systematic error sources, including the dipole (and higher odd
        multipoles of) $E$-field, the Earth's gravitation field and some
        additional Geometrical phases (\Cref{sec:geom_phase_add}).

  \item \emph{Quadrupole polarity switching}. As mentioned in
        \Cref{sec:data_combination}, flipping the polarity of the magnetic
        quadrupoles effectively phase shifts the beta functions. Therefore, a
        significant amount of systematic error  sources that depend on local
        values of the beta function are suppressed.

  \item \emph{Beam splitting}. Applying radial and vertical magnetic fields
        $B_{x,y}$ to split the CR beams enhances the effect of local (skew) quadrupole
        $E$ fields (\Cref{sec:spinbasedalignment}). Splitting the CR
        beams increases local beam offsets that will greatly amplify effects of
        quadrupole and higher order $E$ fields. With such amplification, it is
        possible to measure and control high order $E$ fields via SBA.

  \item \emph{Positive and negative helicities}.  Probing EDM. In addition to CR beam storage,
        bunches with opposite helicities are present too, reserved for polarimeter related systematics.

  \item \emph{Radially polarized bunches}.           Probing DM/DE. Radially polarized            bunches are the
        most sensitive to Vertical Velocity effect
        (\Cref{sec:vertical_velocity}), and some additional geometrical
        phases (\Cref{sec:geom_phase_add}). Radially polarized bunches are
        needed for SBA and also used for the data combination
        (\Cref{sec:data_combination}).

  \item \emph{Vertically polarized bunches}.         Probing simultaneously EDM and DM/DE. Currently, only       radially and
        longitudinally polarized beams were considered. Utilizing spin
        precession data of vertically polarized beams could be used to further
        mitigate systematic error sources. Vertically polarized bunches are
        sensitive to different $\vec{\Omega}_{a}$ components which could be used
        to isolate the EDM component even better (unused in this study).

  \item \emph{Quadrupole strength variation}. Has been proposed first in
        \cite{haciomeroglu_hybrid_2018}, varying the quadrupole strengths lets
        us extrapolate the effective vertical spin precession rate at an
        infinite quadrupole strength by subsequently increasing the focusing
        gradient $k$, where the beam split is minimal (unused in this study).

  \item \emph{Polarization measurement}. Every few seconds the spin direction
        will be rotated around the vertical axis in one direction and
        immediately in the opposite one, in order to have an accurate
        measurement of the beam polarization value as well as of the vertical
        spin component as a function of time. This technique is implicitly
        assumed in this work.
\end{itemize}
\subsection{Conclusion}
The most important systematic error sources in the storage ring proton EDM
experiment are covered. Overall, we have shown that for the specified ring
alignment requirements, the most significant systematic error sources are well below
the target EDM sensitivity. This paper has introduced novel methods of improving
the sensitivity of the experiment such as Symmetric-Hybrid ring design, hybrid
sextupoles for increased SCT, and Spin-based alignment. Combined
with~\cite{anastassopoulos_storage_2016,haciomeroglu_hybrid_2018,edm_proposal}
this work aims to be the constitutive basis for the conceptual and technical design reports (CDR and TDR). We expect to write a white paper as part of the current community effort to evaluate its options and set the priorities for the next five years. Assuming a positive outcome, a CDR is in order, with a parallel effort of a ``string test'' including all the hardware, plus the injection system, of 1/48th of the ring lattice. The purpose would be to test compatibility and cross interactions, after which we will finalize the hardware specifications and the TDR.

\subsection{Acknowledgments}
This work was supported by IBS-R017-D1 Republic of Korea; U.S.
Department of Energy under Grant Contract DE-SC0012704; U.S. Department of
Energy, Office of Science, Office of Nuclear Physics under contract
DE-AC05-06OR23177; 2019VMA0019 of CAS President's International Fellowship Initiative.

\begin{figure}[tbp]
  \centering
  \includegraphics[width=0.99\linewidth]{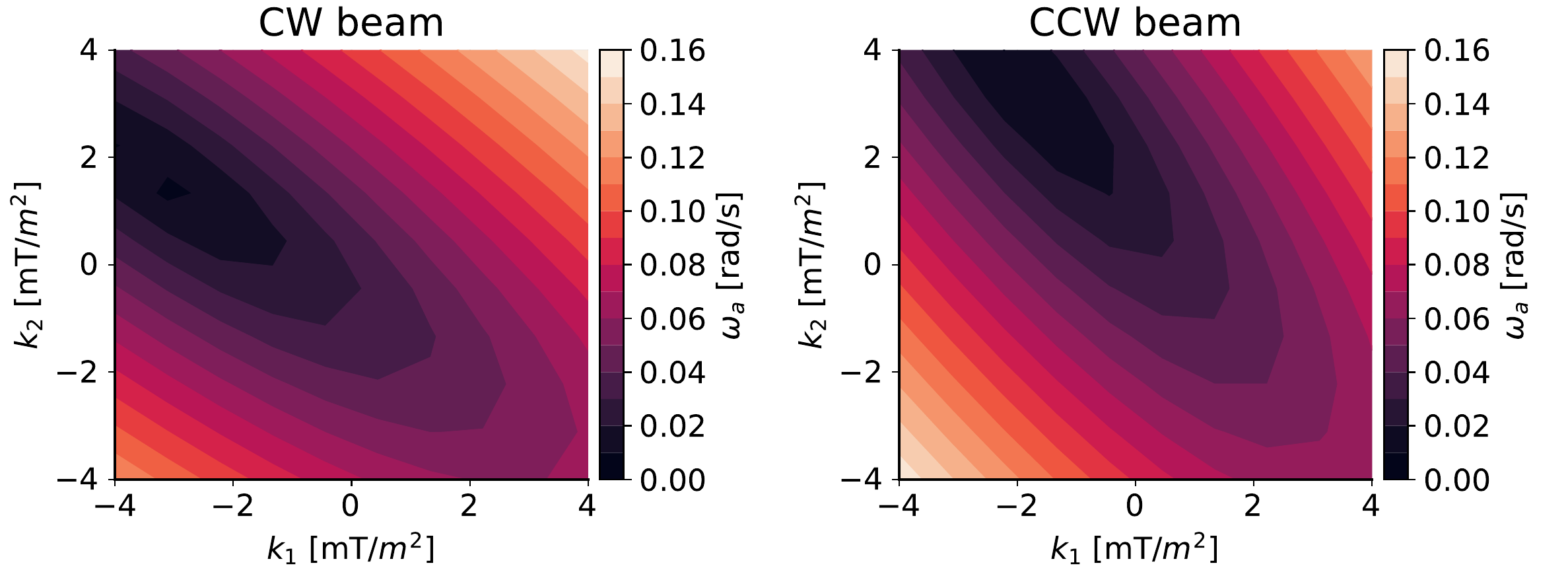}
  \caption{Single particle horizontal precession rate $\omega_{a}$ as a function
    of magnetic sextupole strengths $k_{1},k_{2}$. Darker --- lower $\omega_{a}$; thus, a better SCT. Left --- CW beam, right --- CCW beam. The axis of the symmetry is $k_{1}=-k_{2}$, hence the apparent transposition w.r.t. CR beams.}\label{fig:k1_k2}
\end{figure}
\begin{figure}[tbp]
  \centering
  \includegraphics[width=0.99\linewidth]{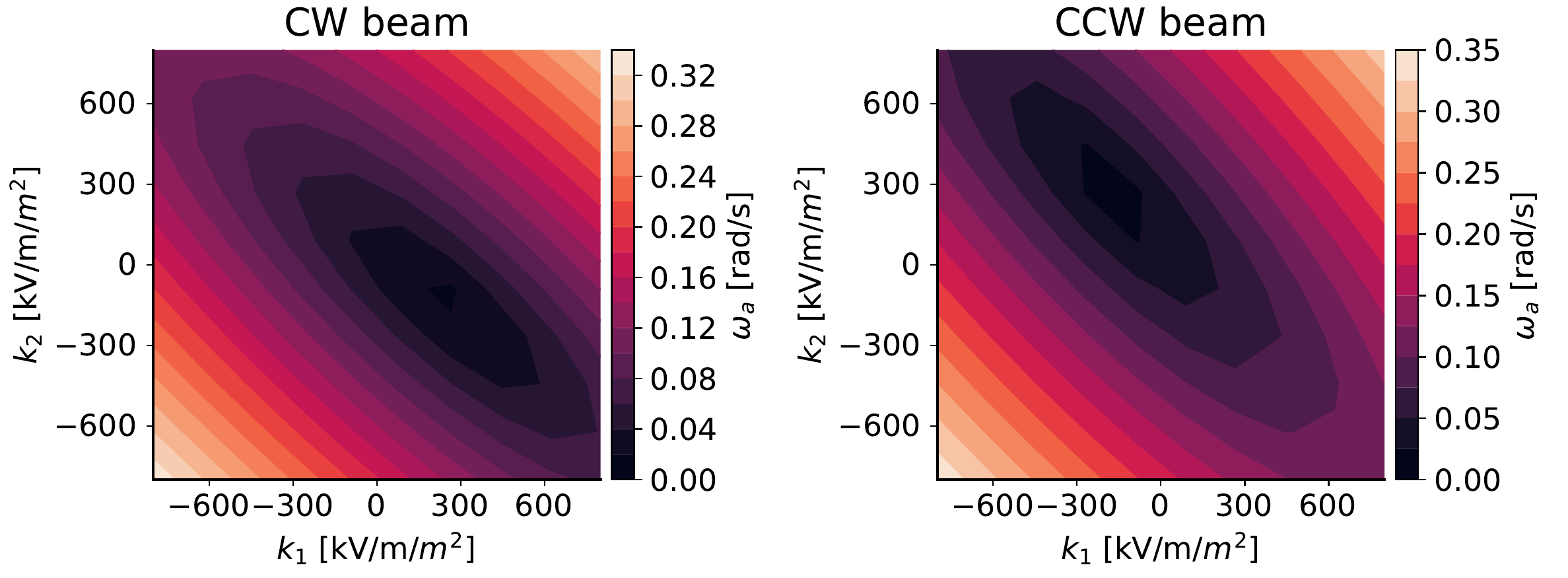}
  \caption{Single particle horizontal precession rate $\omega_{a}$ as a function
    of electric sextupole strengths $k_{1},k_{2}$. Darker --- lower $\omega_{a}$; thus, a better SCT. Left --- CW beam, right --- CCW beam. The axis of the symmetry is $k_{1}=+k_{2}$, hence the apparent transposition w.r.t. CR beams.}\label{fig:elec_k1_k2}
\end{figure}
\begin{figure}[tbp]
  \centering
  \includegraphics[width=0.99\linewidth]{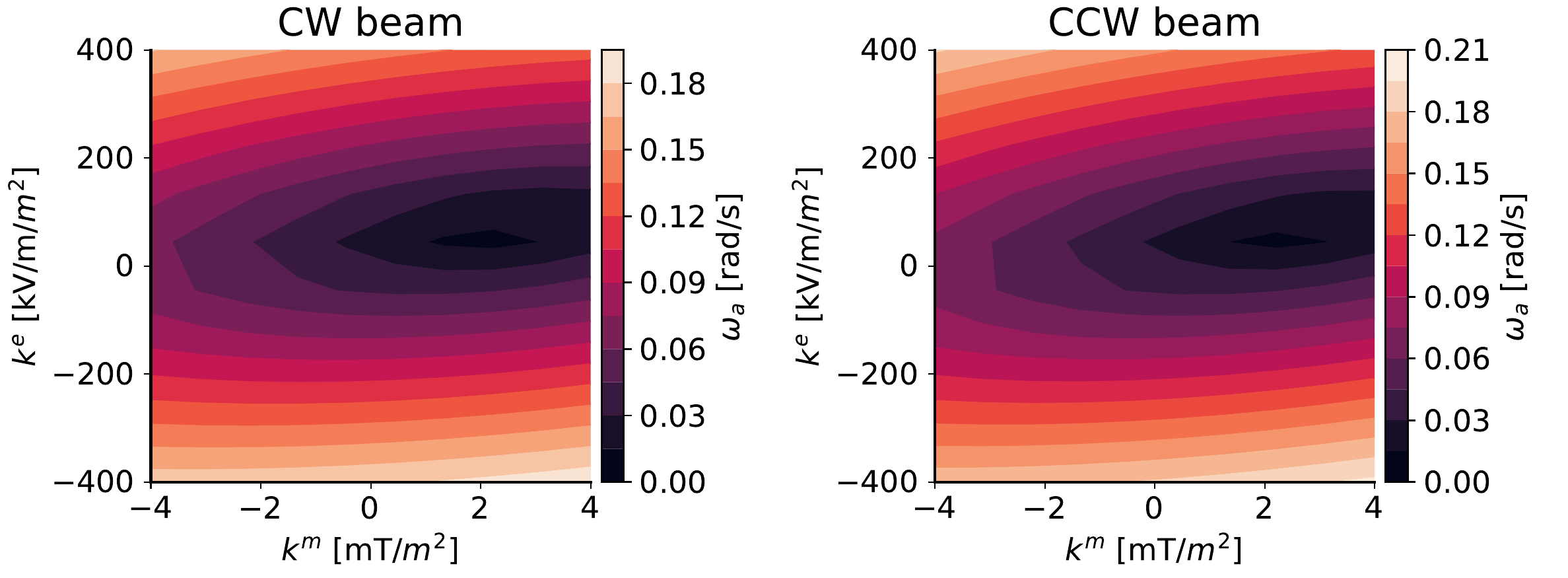}
  \caption{Single particle horizontal precession rate $\omega_{a}$ as a function
    of magnetic and electric sextupole strengths $k^{m},k^{e}$ (\Cref{eq:ke_km}). Darker ---
    lower $\omega_{a}$; thus, better SCT. Left: CW beam, right: CCW beam. The
    effect is perfectly symmetric as the variables can only affect CW-CCW beams
    the same way. Thus, we can improve the SCT for both cases at the same time.}\label{fig:hybrid_k1_k2}
\end{figure}
\begin{figure}[tbp]
  \centering
  \includegraphics[width=0.99\linewidth]{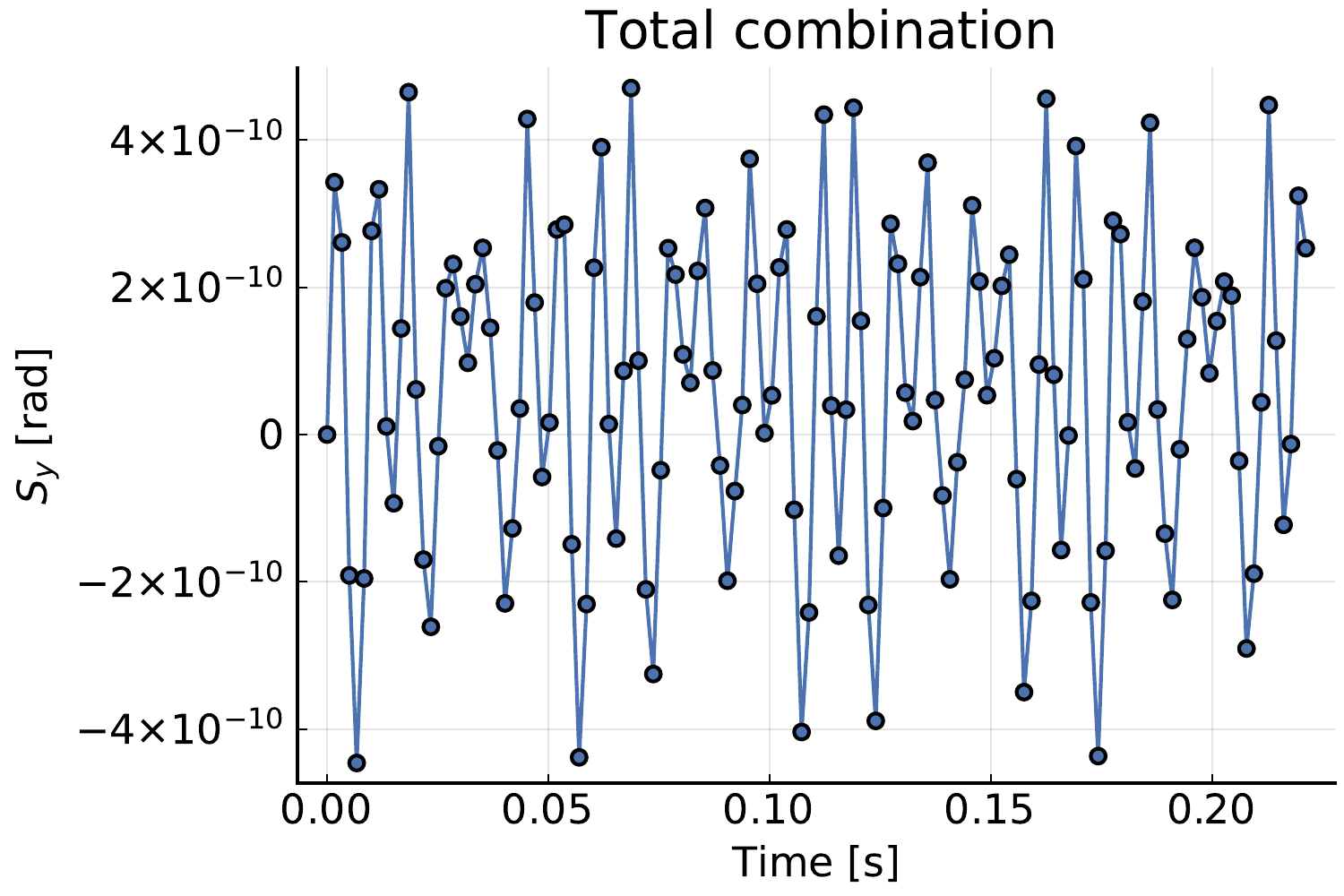}
  \caption{Vertical spin component $S_{y}$ vs. storage time. The signal comes
    from calculation of \Cref{eq:edm_signal,eq:1}, with residual EDM-mimicking
    background precession rate of $dS_{y}/dt < \SI{1}{n rad/s}$ which
    corresponds to the target sensitivity of $d=10^{-29}e\cdot \mathrm{cm}$. The actual
    numerical vertical spin data oscillates rapidly; hence it was arbitrarily averaged onto 40 points.
  }\label{fig:total_precession}
\end{figure}
\appendix

\section{Electrode Material and Design}\label{sec:plates}

\begin{table}[tbp]
  \centering
  \caption{Properties of a single electrode plate}
  \begin{tabular}{ll}
    \toprule
Length                & 104 				cm         \\
Electrode 				Height  & 20.0 				cm        \\
Gap 				Width         & 4.0 				cm         \\
Bending 				Field     & 4.391 				MV/m     \\
Maximum 				Field     & To 				be modelled \\
Voltage 				per Plate & ±87.82 				kV      \\
Bending 				Radius    & 95.49 				m \\
    \bottomrule
\end{tabular}
  \label{tab:electrode}

\end{table}

The ring design imposes strict requirements on the choice of the electrode
material. The electrodes must be compatible with bake-out at $200^{\circ}$C (due
to required vacuum of $\SI{e-10}{Torr}$) and nonmagnetic since the background
magnetic field must be very small ($<\SI{1}{nT}$). Other requirements relate to
having an electrode made of a material that is easy to machine and polish. The
electrode must also be made to very good tolerances to meet the required
alignment and be from a light weight material. From the many choices
considered~\cite{electrode1,electrode2}: stainless steel 316L, niobium,
molybdenum, titanium and TiN-coated aluminum, TiN-coated aluminum shows great
promise.

The studies done on TiN-coated small electrodes~\cite{electrode2} where the aluminum
electrodes were manufactured from Al6061 alloy, required only hours of
mechanical polishing using silicon carbide paper. The coating was about
$\SI{2.5}{\micro \metre}$ thick and the electrodes were baked at $200^{\circ}$C
for 30 hours and achieved $\SI{e-11}{Torr}$. However, these were small
electrodes and coating large pieces will be a challenge. Although the tested
small pieces did not show field emission up to $\SI{14}{MV/m}$, it is known that
large pieces will not perform as well as small pieces.

\Cref{tab:electrode} shows the properties of a single electrode plate. The ring
will use 1152 such plates. The transverse edges of these plates can be shaped
with Rogowski
edge profile and electrostatic modelled to find the maximum field strength. The
model will also speak to field strength everywhere inside the vacuum pipe and
not just between the plates. To bias the plates and support them inside the vacuum
pipe, one can use inverted insulators.  R24 insulators should work as they are
relatively compact, about 10 cm long, and rated to nearly $\SI{200}{kV}$.  Each plate
will be supported by two insulators.  Since these insulators are sold to
medical x-ray community, they should be relatively inexpensive. There are
industry standard cables with R24 connectors. Small alumina insulator spacers
can be used to hold the plates apart, and provide a relatively accurate gap.
These spacers would also minimize bowing due to electrostatic forces. However,
the design should prevent line-of-sight, to avoid the possibility of charging-up
these spacers, and confirm that these small insulator spacers will not distort
the field homogeneity. The spacers would simplify construction and minimize
expense of in-situ alignment. All triple-point junctions (wherever metal touches
insulator in vacuum) could be a source of high voltage breakdown and requires
careful design and shielding\cite{electrode3}.

For each electrode with dimensions roughly $\SI{104}{cm}$ long by $\SI{20}{cm}$
tall by $\SI{4}{cm}$ thick, the total volume is $\SI{8320}{cm}^{3}$.  Given
aluminum density of $\SI{2.7}{g/cm}^{3}$, each electrode weighs roughly $\SI{22}{kg}$ (or 50 pounds). To reduce weight and cost, hollow aluminum electrodes should be considered. One way to mount the electrodes inside the beam pipe is to have the insulators oriented vertical, supporting the plates from below, with gravity helping. However, this breaks the vertical symmetry and may introduce a vertical electric field component. In this case, the electrodes should be mounted from the back side.

To apply high voltage of each polarity, the plates can be daisy-chained together with one supply biasing many plates. Additionally, the high voltage system should be configured such that plates can be biased separately, and negatively with the other plate grounded, to allow gas conditioning when field emission is observed. Effects of high voltage stability and ripple should be investigated.

Since the vertical electric field, $E_{y}$, and its stability are the main systematic uncertainties in the experiment, one must consider very carefully the mechanical alignment of the electrode plates, the flatness of the aluminum plates, polishing and coating, especially such  large pieces, and electrode plates parallelism. In principle, not only the plates but also the insulators, with such a large number of insulators needed, the required tolerances on manufacturing of these has to be also specified.

\section{Extraction of the linear rate from a noisy curve}\label{sec:data}
\begin{figure}[htbp]
  \centering
  \includegraphics[width=0.99\linewidth]{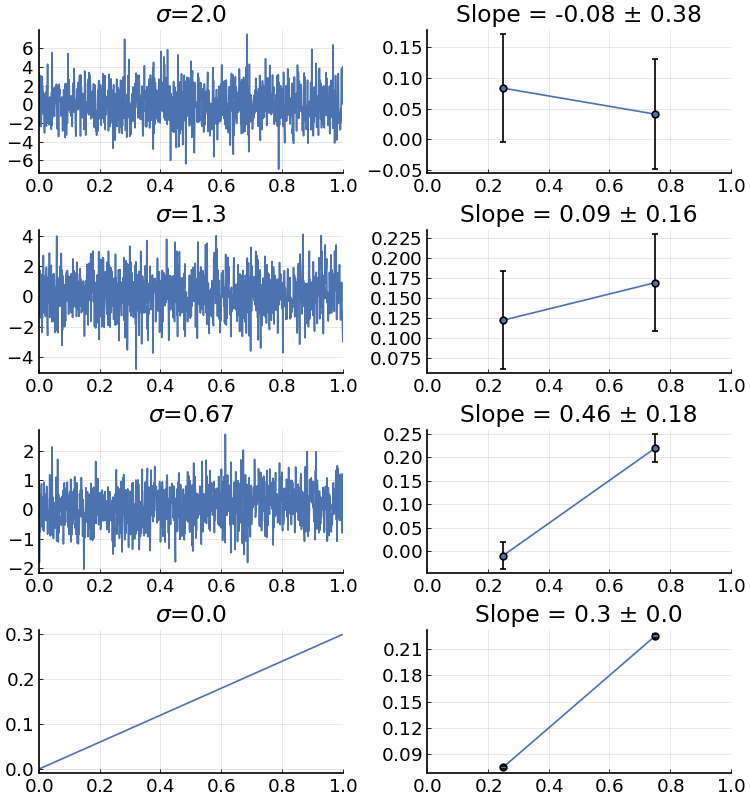}
  \caption{Left column: Random data with an increasing amount of normally
  distributed noise with $\sigma$. The underlying data (last row) has no noise
  ($\sigma = 0$) width with its true slope apparent. With increasing noise levels
  (first 3 rows) the same slope becomes hidden due to large noise. The data contain
  1000 points.\\
  Right column: the data from the left column is averaged into two bins with on
  standard deviation / $\sqrt{n}$ included as the error bars.}
  \label{fig:random_line}
\end{figure}

Measuring the rate (slope) of a linear curve under random noise can only be
done up to some certain statistical significance. For example, the random linear
$x-y$ curve given in \Cref{fig:random_line} (bottom left column) is subjected to random
normal increasing-in-magnitude noise (mid-top left column). Although there are a few
possibilities to extract the underlying rate (slope), in this work the data is
averaged onto two points with the standard deviations noted ---
\Cref{fig:random_line} (right column).  Next, a least squares fit is performed to
assess the slope and the standard error of the slope. 

With the obtained estimate of the underlying rate, analogous to estimation of the
vertical spin precession rate, a confidence interval is reported as the maximal
possible rate. Due to such statistical uncertainty (not numerical uncertainty),
low precession rate values happen to be irregular ---
\Cref{fig:quad-misalignment-both,fig:e_multipoles,fig:b_multipoles}. The simulation is unable to achieve times of order of
$~\SI{1000}{s}$ (real experiment) under reasonable time and precision
requirements. Hence, it has been performed such that the random noise in
vertical spin component due to free betatron and spin oscillations does not
contribute more than the target precision level of $\SI{e-9}{rad/s}$. 

\section{Data combination}\label{sec:data_combination}
Out-of-plane precession direction due to
a genuine EDM signal would be opposite for positive helicity CR beams in
the same storage ring. Namely \footnote{$y$ is vertical in the lab frame.},
\begin{dmath}
  \qty(\frac{dS_{y}}{dt})_{\textnormal{EDM}} = \frac{1}{2}\qty(\frac{dS_{y}}{dt})_{\textnormal{CW}} - \frac{1}{2}\qty(\frac{dS_{y}}{dt})_{\textnormal{CCW}}
  \label{eq:edm_signal_cwccw}
\end{dmath}
vertical spin precession due to the EDM is the difference of the vertical
precessions of CW and CCW beams, with a factor of
$1/2$ to compensate for the doubling of the true EDM signal.

Complementing the simultaneous storage of CR beams, we can
have another symmetrical flip --- Polarity
switch. It is the act of switching the direction of the
currents in the magnetic quadrupoles. Quadrupoles being maximally current
dominated makes it possible to phase shift the lattice beta functions by
reversing direction of the currents in all the magnetic quadrupoles.
Then, the EDM signal is given as,
\begin{dmath}
  \qty(\frac{dS_{y}}{dt})_{\textnormal{EDM}} = \left[ \frac{1}{4}\qty(\frac{dS_{y}}{dt})_{\textnormal{CW}} - \frac{1}{4}\qty(\frac{dS_{y}}{dt})_{\textnormal{CCW}} \right]_{\textnormal{Polarity 1}}
  + \left[ \frac{1}{4}\qty(\frac{dS_{y}}{dt})_{\textnormal{CW}} - \frac{1}{4}\qty(\frac{dS_{y}}{dt})_{\textnormal{CCW}} \right]_{\textnormal{Polarity 2}}.
  \label{eq:edm_signal}
\end{dmath}
In addition to the polarity switch of the quadrupoles, one can choose to
 change the quadrupole gradient --- $k$ to extract
$dS_{y}/dt \propto 1/k$ in the asymptotic
limit of $1/k \rightarrow 0$ as has been first suggested
in~\cite{haciomeroglu_hybrid_2018}.

In addition to the vertical spin precession, the spin would also
(inevitably) precess into radial direction, assuming the lattice conditions
listed in \Cref{sec:realistic}. Radial spin precession could
 create a vertical precession
(\Cref{sec:vertical_velocity}).

We can model this case analytically and compensate for such radial
spin precession. We denote the vertical spin precession rate as the combined
effect from both radial and longitudinal polarizations,
\begin{equation}
\frac{dS_y}{dt} = \eta' S_s + \delta' S_x. \label{eq:c3}
\end{equation}
\(\eta'\) indicates EDM-like precession that only
happens due to a longitudinal
spin component \(S_s\) and \(\delta'\) indicates dark matter-like precession that
happens due to a radial spin component $S_{x}$. \(\eta'\) and \(\delta'\) show
only the combined background effect. For example, \(\delta'\) directly
contains the vertical velocity and other systematics that only
happen when the spin is radial.

Assuming that an initially longitudinally polarized bunch precesses into radial
direction linearly with time,
\begin{equation}
    \frac{dS_x}{dt} = \Gamma S_s, \label{eq:c4}
\end{equation}
where \(\Gamma\) stands for \((g-2)\) -- like in-plane spin precession.

\begin{figure}[tbp]
  \centering
  \includegraphics[width=0.99\linewidth]{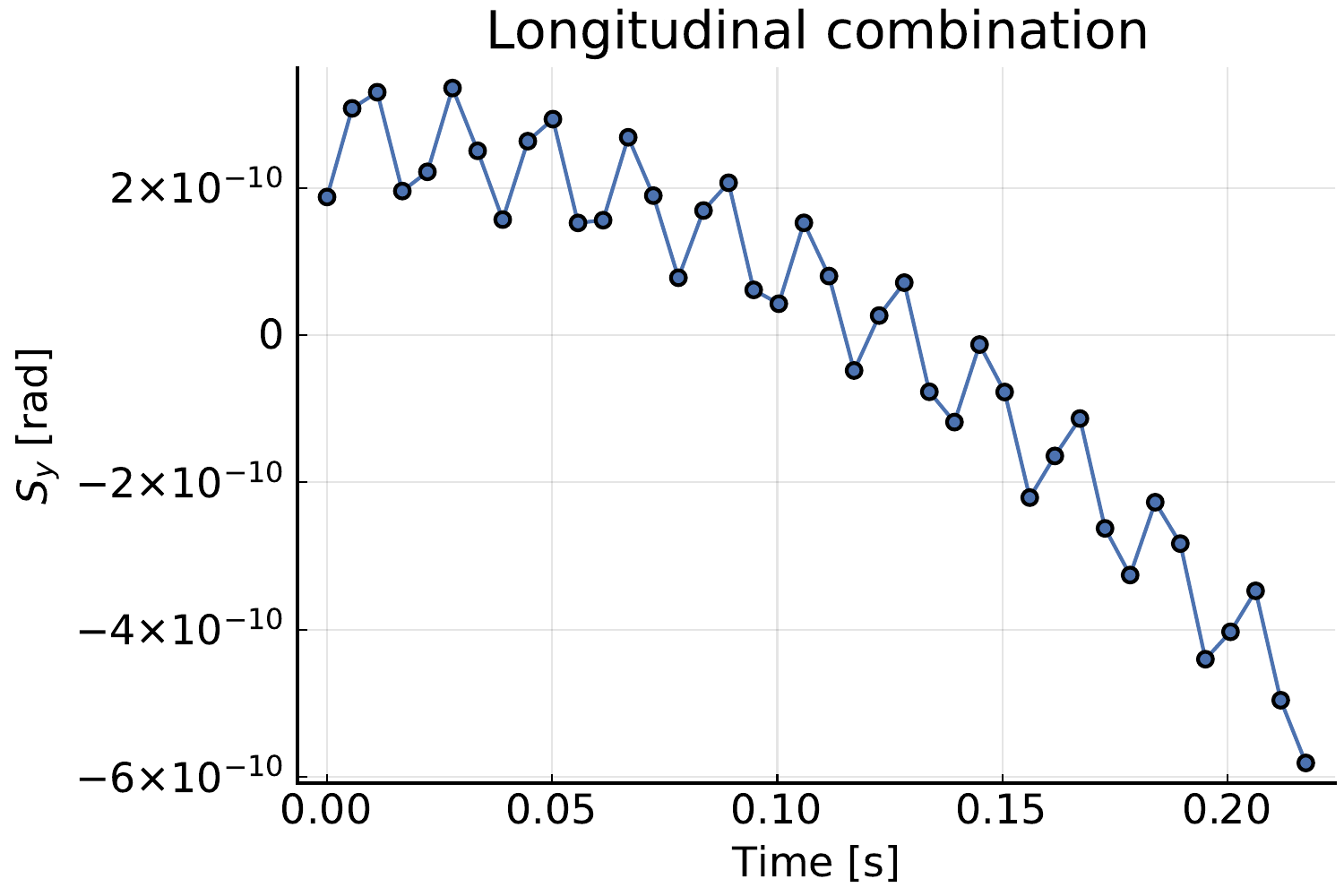}
  \caption{Effective vertical spin component precession data of CW-CCW and
    both quadrupole polarities calculated from \Cref{eq:edm_signal}. The
    quadratic behavior of this curve is explained by \Cref{eq:1}.
  }\label{fig:long_combination}
\end{figure}

If the radial and vertical spin precession rates are small, i.e.
\(1\approx S_s \gg S_x, S_y\) at all
times, the coupled differential equations (\Cref{eq:c3,eq:c4}) have the solution,
\begin{equation}
\label{eq:1}
S_s\eta' = \frac{d}{dt} \qty(S_y - \delta' \Gamma t^2 S_s / 2).
\end{equation}
The quadratic in time behavior caused by the drift into radial spin
direction is confirmed --- \Cref{fig:long_combination}.

\begin{figure}[tbp]
  \centering
  \includegraphics[width=0.99\linewidth]{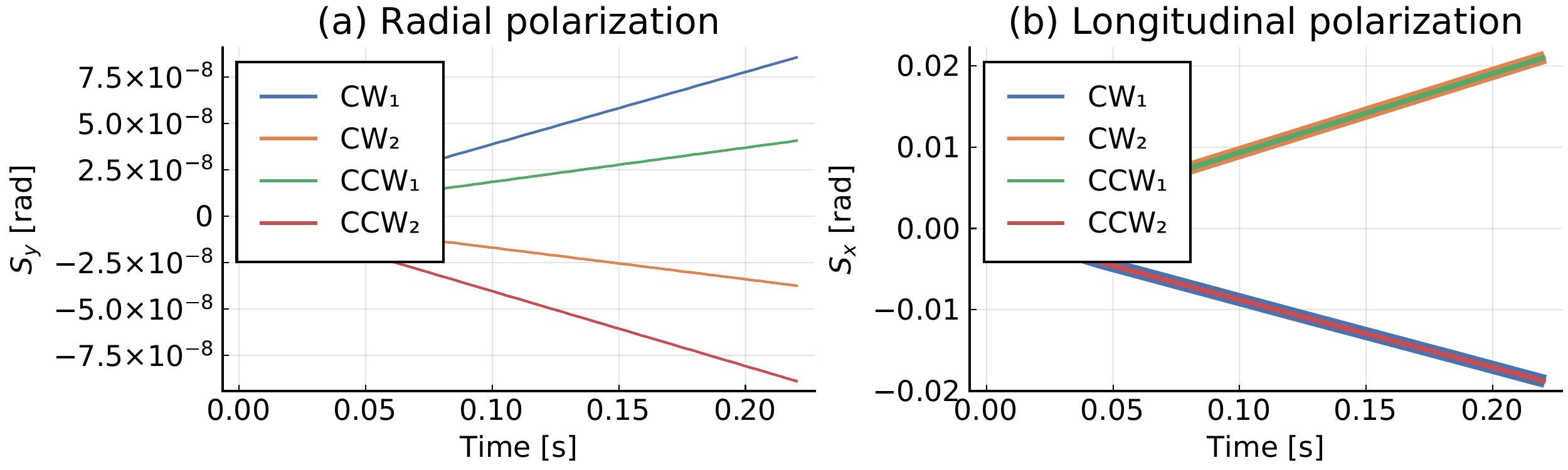}
  \caption{(a): Vertical spin precession of a radially polarized beam for all  beam
    directions and polarities (subscripts). Main contribution of the spin growth originates from Vertical Velocity effect \Cref{sec:vertical_velocity}.
    (b): Radial spin precession of a longitudinally polarized beam for all beam
    directions and polarities (subscripts). The spin precesses into radial
    direction due to RF frequency mismatch and non-zero average $B_{y}$ present
    in the storage ring due the misalignment of magnetic quadrupoles.
  }\label{fig:rad_long}
\end{figure}

First, $\delta'$ and $\Gamma$ are
estimated from spin precession data --- \Cref{fig:rad_long}. After
redefining the effective $\frac{dS_{y}}{dt}$ as,
\[\frac{dS_{y}}{dt} \gets \frac{d}{dt} \qty(S_y - \delta' \Gamma t^2 S_s / 2)\]
$\eta'$ can now be correctly estimated with \Cref{eq:1,eq:edm_signal}.
The total combination result is given on \Cref{fig:total_precession}, from
which it is clear that $\eta' < \SI{1}{nrad/s}$.

\section{High precision tracking}\label{sec:tracking}
The Lorentz equation governs the dynamics of a particle in EM fields,
\begin{equation*}
  \frac{d\vec{\beta}}{dt} = \frac{q}{m\gamma c} \left[ \vec{E} + c \vec{\beta} \times \vec{B} - \beta(\vec{\beta} \cdot \vec{E})    \right].
  \label{eq:beam-dynamics}
\end{equation*}
However, its perturbative expansion in particle optical coordinates is more
practical for storage rings and accelerators, as we are using natural variables
of interest ~\cite{berz_introduction_2014},
\begin{align*}
  x' &= a \qty(1 + hx) \frac{p_{0}}{p_{s}} \\
  a' &= \qty(1 + hx) \qty[\frac{\gamma}{\gamma_{0}}\frac{E_{x}}{\chi_{e0}}\frac{p_{0}}{p_{s}} + b\frac{B_{s}}{\chi_{m0}}\frac{p_{0}}{p_{s}} - \frac{B_{y}}{\chi_{m0}}] + h \frac{p_{s}}{p_{0}}\\
  y' &= b \qty(1 + hx) \frac{p_{o}}{p_{s}} \\
  b' &= \qty(1 + hx) \qty[\frac{\gamma}{\gamma_{0}}\frac{E_{y}}{\chi_{e0}}\frac{p_{0}}{p_{s}} + \frac{B_{x}}{\chi_{m0}} - a\frac{B_{s}}{\chi_{m0}}\frac{p_0}{p_s}],
\end{align*}
where prime indicates differentiation with respect to $s$ (beamline travel
distance) and subscript $0$ is the quantity with respect to the reference particle.
In this curvilinear (Frenet--Serret) coordinate system, $x$ indicates radial
deviation from the reference orbit, $y$ indicates vertical deviation, and $s$
points along the direction of motion of the reference particle. Hence, the
momentum in this coordinate system is measured in
$\vec{p}/p_{0} = (a, b, p_{s}/p_{0})$. In other variables, $h=1/R_{0}$ indicates
curvature for the reference orbit and $\chi_{e0}, \chi_{m0} = p_{0}v/Ze$, $p_{0}/Ze$
the electric and magnetic rigidities.

The spin vector should then be integrated with the T-BMT equation~\cite{bmt,bmt2}
given in Cartesian coordinates as follows,
\begin{equation*}
    \frac{d \vec{S}}{dt} = \vec{\Omega}\times\vec{S}
\end{equation*}
\begin{dmath}
  \frac{d \vec{S}}{dt} = \frac{q}{m}\vec{S}\times
  \left[
    \left(a+\frac{1}{\gamma}\right) \vec{B} - \frac{a\gamma}{\gamma+1} \vec{\beta} (\vec{\beta} \cdot \vec{B}) - \left(a+\frac{1}{\gamma+1} \right) \frac{\vec{\beta} \times \vec{E}}{c}
    +
    \frac{\eta}{2} \left(\frac{\vec{E}}{c} - \frac{\gamma}{\gamma+1}\frac{\vec{\beta}}{c}\qty(\vec{\beta}\cdot\vec{E})+\vec{\beta}\cross\vec{B}\right)
  \right].
  \label{eq:spin-dynamics}
\end{dmath}

However, the spin normalized to unity measured in terms
of $\vec{S} = (S_{x}, S_{y}, S_{s})$ --- radial, vertical, and longitudinal spin
components, the original T-BMT equation (\Cref{eq:spin-dynamics}) needs to be
modified as,
\begin{equation}
  \vec{S}' = \qty(\vec{\Omega}t' - h\hat{y}) \times \vec{S},
\end{equation}
in order to compensate for the rotation of the coordinate system itself, and
take into account that we want the derivative w.r.t. to $s$ integration variable
(here $\hat{y}$ refers to vertical---out-of-the-plane --- unit vector).

Each of the storage ring elements have been tracked separately in order to avoid
discontinuities in EM fields that directly lead to unstable numerical
integration. Electric bending plates were hard-edge approximated. The fields
inside cylindrical deflectors with a focusing index $n=m+1$ are given
as~\cite{metodiev_fringe_2014,metodiev_analytical_2015},
\begin{align*}
  E_x &= - E_0  \qty(1 - \frac{nx}{R_0} + \frac{n(n+1)x^2}{2R_0^2})\\
  E_y &= - E_0  \qty((n-1)  \frac{y}{R_0}).
\end{align*}
It is important to note that in order to meet the precision requirements, second
order terms ($x^{2}$) must be considered to have precise spin integration.

\section{Detailed analysis of the Geometrical Phases}\label{sec:geom_phase_add}
The terms in the T-BMT equation proportional to the \(E\)
field are the main mechanism for systematic error sources, as \(B\) fields are
naturally shielded by the quadrupoles. In addition to the dipole \(E\) field,
vertical velocity, and quadrupole \(E\)
field (\Cref{sec:dipole_e,sec:spinbasedalignment}) there exist additional ways
of creating background vertical spin precession.

The list of the possible terms from \Cref{eq:spin-dynamics} is given
below (in the order of importance attributed by authors),
\begin{enumerate}
\item \(dS_y/dt \propto S_x \cdot \beta_y \cdot E_x\), discussed in \Cref{sec:vertical_velocity}, also recognized as ``twist'' distortion.
\item \(dS_y/dt \propto S_s \cdot \beta_s \cdot E_y\), discussed in \Cref{sec:dipole_e}
\item \(dS_y/dt \propto S_s \cdot \beta_y \cdot E_s\), will receive additional
treatment in this section.
\item \(dS_y/dt \propto S_x \cdot \beta_x \cdot E_y\), will receive additional
treatment in this section.
\end{enumerate}
The \(dS_y/dt \propto S_s \cdot \beta_y \cdot E_s\) term directly couples to the
longitudinal polarization (EDM search), thus circumventing
its effect via SBA is not trivial. As it has been argued in
\Cref{sec:vertical_velocity},
the average effect of the \(S_s \cdot \beta_y \cdot E_s\) term would
not be zero. Nevertheless, from a energy conservation stand point,  \(\int E_s
ds = 0\) per each deflector. A static electric field cannot
induce a net acceleration (or deceleration) on a passing particle. Hence,
the effect of this term effectively applies to non-static longitudinal electric
fields. High-precision numerical spin tracking has shown that even in the case where
\(\int E_{s} ds \ne 0\), for example, due a changing-in-time magnetic flux
through the storage ring plane, the resulting false EDM signal is below
$\SI{1}{nrad/s}$ for fields \(E_{s}<\SI{5}{V/m}\) normally distributed along
the ring azimuth.

The \(dS_y/dt \propto S_x \cdot \beta_x \cdot E_y\) term couples to the radial
polarization, hence the effect is suppressed via SBA (\Cref{sec:spinbasedalignment}).
Isolation of this effect is challenging, as its contribution is orders of magnitude below the vertical
velocity effect. A vertical velocity is inadvertently created when probing for the
\(S_x \cdot \beta_x \cdot E_y\) effect,
as \(E_y\ne 0\) fields create \(\beta_y\ne 0\). This effect is negligible for
the current sensitivity goals and could be ignored for all practical purposes.

\section{Optimum sextupole strength search}\label{sec:optimum_sextupole}
In \Cref{fig:elec_k1_k2,fig:k1_k2,fig:hybrid_k1_k2} the optimal
sextupole strengths pair is obtained by a rough 2-dimensional parameter
sweep, followed by numerical optimization to find the finer minimum.
It is worthwhile to show that finding the optimal pairs for magnetic and electric
sextupoles separately is sufficient to infer the value for the optimum strength
needed for hybrid sextupoles.

Let us suppose that $k_{1}^{m}=\alpha_{1},k_{2}^{m}=-\beta_{1}$ is the optimal
pair for magnetic sextupoles for CW beam, with
$k_{1}^{m}=\beta_{1},k_{2}^{m}=-\alpha_{1}$ for the CCW case. Similarly,
$k_{1}^{e}=-\alpha_{2},k_{2}^{e}=\beta_{2}$ and
$k_{1}^{e}=\beta_{2},k_{2}^{e}=-\alpha_{2}$ for CW and CCW directions
respectively with electric sextupoles.

By observing the symmetry in \Cref{fig:elec_k1_k2,fig:k1_k2}, we can infer that,
\begin{align}
  k_{1}^{m} &= -\frac{\alpha_{1}}{\alpha_{2}} \times k_{1}^{e} & k_{2}^{m} &= -\frac{\beta_{1}}{\beta_{2}}\times k_{2}^{e} \qq{for CW} \\
  k_{1}^{m} &= \frac{\beta_{1}}{\beta_{2}} \times k_{1}^{e} & k_{2}^{m} &= \frac{\alpha_{1}}{\alpha_{2}}\times k_{2}^{e} \qq{for CCW}
\end{align}
i.e. only a sign change is required for the transition from electric to magnetic or
vice versa. We can also infer the conversion factor from electric to magnetic
sextupoles. Following the lines of the symmetry, we can find the optimal pair
for the hybrid sextupoles case in magnetic units,
\begin{align}
  M_{1} + E_{1} &= \alpha_{1} & -M_{1} + E_{1} = -\beta_{1}
\end{align}
solving for each $M_{1}=\qty(\alpha_{1} + \beta_{1})/2$ and
$E_{1}=\alpha_{2}/\alpha_{1} \times \qty(\alpha_{1} - \beta_{1})/2$ we get the
optimal pair for each case in proper units.
\Cref{fig:elec_k1_k2,fig:k1_k2,fig:hybrid_k1_k2} verify these analytical
estimations to 1\% accuracy.

\bibliography{refs}

\end{document}